\documentclass{article}

\usepackage{graphicx}
\usepackage{amsmath}
\usepackage{algorithm2e}
\usepackage{multicol}
\usepackage{nomencl}
\usepackage{ifthen}
\usepackage{float}
\usepackage{placeins}

\usepackage{cleveref} 

\makenomenclature   

\renewcommand{\nomgroup}[1]{ %
  \ifthenelse{\equal{#1}{D}} %
             {\item[\textbf{Design}]} { %
  \ifthenelse{\equal{#1}{T}} %
             {\item[\textbf{Statistics}]} { %
  \ifthenelse{\equal{#1}{O}} %
             {\item[\textbf{Operators}]} { %
  \ifthenelse{\equal{#1}{E}} %
             {\item[\textbf{Sets and Cardinality}]} { %
  \ifthenelse{\equal{#1}{S}} %
             {\item[\textbf{Sub- and Super-scripts}]} }}}}}

\usepackage{amsmath}  
\usepackage{amsfonts} 
\usepackage{scalerel,stackengine} 
\usepackage{graphicx} 
\usepackage{caption}  
\usepackage{mathtools}
\usepackage{forest}   

\def\R#1{\mathbb{R}^{#1}}

\def\N#1{\mathbb{N}^{#1}}

\newcommand{\vsym}[1]{\boldsymbol{#1}}
\def\v#1{\vsym{#1}} 

\newcommand{\vd}{\boldsymbol{d}}

\newcommand{\mT}{\boldsymbol{T}}
\newcommand{\mU}{\boldsymbol{U}}

\newcommand{\mX}{\boldsymbol{X}}

\newcommand{\cC}{\mathcal{C}}

\newcommand{\cF}{\mathcal{F}}

\newcommand{\cP}{\mathcal{P}}

\newcommand{\cR}{\mathcal{R}}

\newcommand{\mrm}{\mathrm}
\def\E{\mathbb{E}} 

\def\P{\mathbb{P}} 

\def\Cov{\mrm{Cov}} 

\def\V{\mathrm{V}} 
\newcommand{\dN}{\mathcal{N}} 

\def\i1{\mathbb{1}}

\stackMath
\newcommand\reallywidehat[1]{%
\savestack{\tmpbox}{\stretchto{%
  \scaleto{%
    \scalerel*[\widthof{\ensuremath{#1}}]{\kern-.6pt\bigwedge\kern-.6pt}%
    {\rule[-\textheight/2]{1ex}{\textheight}}
  }{\textheight}%
}{0.5ex}}%
\stackon[1pt]{#1}{\tmpbox}%
}

\newcommand{\ftree}[1]{%
\begin{forest}
for tree={
    font=\ttfamily,
    grow'=0,
    child anchor=west,
    parent anchor=south,
    anchor=west,
    calign=first,
    edge path={
      \noexpand\path [draw, \forestoption{edge}]
      (!u.south west) +(7.5pt,0) |- node[fill,inner sep=1.25pt] {} (.child anchor)\forestoption{edge label};
    },
    before typesetting nodes={
      if n=1
        {insert before={[,phantom]}}
        {}
    },
    fit=band,
    before computing xy={l=15pt},
  }%
  #1%
\end{forest}
}

\usepackage{numdef}   
\num
\num
\num
\num
\num
\num
\num
\num

\num
\num
\num
\num
\num
\num
\num
\num

\num
\num
\num
\num
\num
\num
\num
\num

\num
\num
\num
\num
\num
\num
\num
\num

\num
\num
\num
\num
\num
\num
\num
\num

\num
\num
\num
\num
\num
\num
\num
\num

\num
\num
\num
\num
\num
\num
\num
\num

\num
\num
\num
\num
\num
\num
\num
\num

\num
\num
\num
\num
\num
\num
\num
\num

\num
\num
\num

\newcommand{\margin}{precision margin}
\newcommand{\Margin}{Precision margin}
\newcommand{\MARGIN}{Precision Margin}
\newcommand{\TLA}{PM}

\providecommand{\e}[1]{\ensuremath{\times 10^{#1}}}

\graphicspath{{./}}

\usepackage[margin=1.0in]{geometry}

\begin{document}

\author{Zachary del Rosario, Richard W. Fenrich, and Gianluca Iaccarino}




\title{Cutting the Double Loop:\\ Theory and Algorithms for Reliability-Based
  Design Optimization with Statistical Uncertainty}

\maketitle

\abstract{Statistical uncertainties complicate engineering design --
  confounding regulated design approaches, and degrading the performance of
  reliability efforts. The simplest means to tackle this uncertainty is
  \textbf{double loop simulation}; a nested Monte Carlo method that, for
  practical problems, is intractable. In this work, we introduce a flexible,
  general approximation technique that obviates the double loop. This
  approximation is constructed in the context of a novel theory of reliability
  design under statistical uncertainty: We introduce metrics for measuring the
  efficacy of RBDO strategies (\textbf{effective margin} and \textbf{effective
    reliability}), minimal conditions for controlling uncertain reliability
  (\textbf{precision margin}), and stricter conditions that guarantee the
  desired reliability at a designed confidence level. We provide a number of
  examples with open-source code to demonstrate our approaches in a reproducible
  fashion.}


\section{Introduction} \label{sec:introduction}
Uncertainty complicates design. Unknown loads motivate safety factors;
manufacturing fluctuations motivate material property knockdowns. When
uncertainty is modeled by a random variable, additional uncertainty arises when
fitted distribution parameters are estimated from data, leading to statistical
uncertainty.

Statistical uncertainty represents a lack of knowledge in a system or design,
with the potential for improvement in performance or safety. Such uncertainty
can lead to degraded performance; for example, Park et al.\cite{park2014coupon}
demonstrated significant weight penalties due to sampling uncertainties in
coupon and element testing. Gains in engineering design can be made through the
acquisition of more information, though the question remains of how to
confidently and efficiently guarantee the reliability of a system's performance
and safety under statistical uncertainty.

Optimizing system performance while constrained by failure probability goes by
the name \textbf{reliability based design optimization} (RBDO). When statistical
uncertainties are modeled as parameters to input distributions, they induce
second-order uncertainties similar to a hierarchical
model.\cite{gelman2013bayesian} These uncertainties are most simply handled
through a double loop Monte Carlo simulation\cite{ito2018conservative} over a
sampling distribution or hyperprior. Of course, this approach is multiplicative
in its expense, rendering all but the simplest problems intractable.

Further, in reviewing the literature it was unclear to us how to \emph{measure}
the effects of statistical uncertainties in RBDO, let alone how to control
realized design
reliability.\cite{ito2018conservative,der2008analysis,noh2009reliability} The
aforementioned works introduce approaches that are distinct in how they
introduce engineering conservatism, but are similar in that they recover the the
`true' reliability in the case of perfect information. This is in contrast with
other design practices outside the framework of RBDO, such as those utilizing
\emph{basis values}. Further, in statistical inference, there exists the notion
of confidence intervals, which guarantee frequentist properties of
coverage;\cite{kenett1998} we have not found a similar notion in the context of
RBDO. Our work was in part motivated by a desire for useful theory by which to
compare and contrast different approaches to managing statistical uncertainties.

In this work, we introduce the metrics of \textbf{effective margin} and
\textbf{effective reliability} to assess the performance of RBDO strategies
incorporating statistical uncertainties. To control effective reliability, we
introduce minimum conditions that define \textbf{\margin} (\TLA{}). To show the
concept's generality, we formally prove that the \textbf{conservative
  reliability index} (CRI) of Ito et al.\cite{ito2018conservative} is a \TLA{}.
We also present two implementations of \TLA{}, both carrying unique advantages
and challenges. The second of these implementations -- \textbf{margin in
  probability} (MIP) -- adds \emph{just enough} margin to guarantee the desired
reliability at a known confidence level. We call this property
\textbf{confidently conservative} (C2), and regard it as a translation of
statistical coverage to engineering design practice.

While our examples in this work consider materials characterization, the \TLA{}
concept is flexible enough to apply to any case of sampling uncertainty. The
particular \emph{approximations} of \TLA{} presented in this work are restricted
to cases of modeled randomness, where a specific (analytic) joint PDF is
selected to model variable quantities -- this choice is in line with existing
Department of Defense probabilistic design
methodologies.\cite{mil-hdbk-17-3e1997}

Of course, we are not the first to tackle the double loop issue. Der
Kiureghian\cite{der2008analysis} carried out reliability design over a Bayesian
posterior distribution, effectively incorporating statistical uncertainties into
a single loop; however, his \textbf{predictive reliability index} does not add
any form of margin. Noh et al.\cite{noh2009reliability} tackle the same issue by
perturbing the estimated moments of an normal distribution, assuming that there
exists a transform to standard normal space. Our approach is more general, in
the sense that we work directly in the original probability space of the posed
random variable model. The work of Ito et al.\cite{ito2018conservative} is
closely related to what we suggest, though similarly assumes a transform to
standard normal space, and does not guarantee the C2 property. We draw a close
comparison between their CRI and our proposed MIP approach. Note that some other
authors refer to the form of uncertainty we consider as \textbf{epistemic}, e.g.
Ito et al.\cite{ito2018conservative}. We use a more specific terminology --
statistical uncertainty -- as we do not claim our approach is appropriate for
\emph{all} epistemic uncertainties (such as unknown unknowns), but instead note
that our work addresses many of the same issues commonly referred to as
epistemic uncertainties.

Our approximation technique is a form of Monte Carlo
reweighting,\cite{fonseca2007efficient} but using the likelihood ratio (LR)
gradient estimation technique to approximate parameter gradients at negligible
additional cost.\cite{l1990unified,li2011likelihood} Note that ``double loop''
is sometimes used to refer to a reliability analysis nested within an
optimization loop;\cite{nguyen2010single} we use this term in its other commonly
accepted meaning to refer to nested Monte Carlo.

An outline of this article is as follows: Section \ref{sec:motivation} presents
the motivating issue through a simple structural sizing problem, illustrating
the effects of sampling uncertainty on both standard industry practice and a
`plug-in' RBDO approach. Here we introduce the metrics of effective margin and
effective reliability. Section \ref{sec:sampling-margin} introduces the
\margin{} concept, presents two implementations, and provides comparisons
against the previously-introduced methods. The two implementations apply margin
in either physical or probability space, and present different advantages and
challenges. Section \ref{sec:estimation} provides practical estimation
procedures to enable the computation of \TLA{} -- the techniques introduced here
add negligible computational cost, and are simple to incorporate within an RBDO
framework. Section \ref{sec:demonstration} demonstrates the \TLA{} methodology
on a common RBDO test case, while Section \ref{sec:discussion} retrospects,
providing context and sketching future directions.

Our aim is to constructively comment on the practice of engineering design, and
to illustrate a potential avenue for the continued development of our
profession. In the spirit of facilitating this development, a companion GitHub
repository\footnote{url: https://github.com/zdelrosario/bv-questionable}
contains all the code necessary to generate the results in the present work, and
to serve as a reference implementation for the suggested algorithms.

\section{Motivating Issue} \label{sec:motivation}
We first introduce the design problem of sizing for uniaxial tension, and
formulate the problem in a reliability-based design framework, in order to
illustrate the effects of statistical uncertainty on reliability. We introduce
two families of approaches of dealing with uncertain material properties, first
studying approaches using a basis value, and second directly modeling the
variable material with `plug-in' parameter estimates. We employ all approaches
at different cases of desired reliability, and demonstrate that none produce
desirable results, motivating the introduction of \margin{} in the section to
follow.

\subsection{Uniaxial Tension Sizing}
For illustrative purposes we introduce a structural sizing problem, whose
simplicity highlights the issue of statistical material property uncertainties.
We consider sizing the wall thickness $t$ of a hollow cylinder of given radius
$r$; this has cross sectional area given by $A(t) =
\pi\left((r+t)^2-r^2\right)$. We take the applied tensile force to have a known
distribution $F\sim \dN(\mu_f,\tau^2_f)$, while the material ultimate tensile
strength has a ground truth distribution $U\sim \dN(\mu_{u},\tau^2_{u})$. For
simplicity, we model these variables as independent Gaussians; one could easily
use lognormal variables to enforce positivity, which would not materially change
our conclusions. Table \ref{tab:param-tension-true} summarizes the ground truth
parameter values used in this study.

\nomenclature[D]{$\vd$}{Design Variables}
\nomenclature[D]{$\sigma$}{Stress}
\nomenclature[D]{$f$}{Load}
\nomenclature[T]{$N$}{Normal Distribution}
\nomenclature[T]{$\mu$}{Mean}
\nomenclature[T]{$\tau^2$}{Variance}
\nomenclature[D]{$u$}{Ultimate Strength}

\begin{table}[!ht]
  \centering
  \caption{Ground truth parameters for uniaxial tension example. We assume a
    material coefficient of variation of $10\%$, a high but realistic value for
    advanced composite materials.\cite{oar2003}}
  \label{tab:param-tension-true}
  \begin{tabular}{@{}lll@{}}
    Parameter  & Value & Units \\
    $\mu_{u}$  & $600$ & MPa \\
    $\mu_f$    & $100$ & N \\
    $\tau_{u}$ & $60$ & MPa \\
    $\tau_f$   & $10$  & N
  \end{tabular}
\end{table}

In general, failure of a structure is modeled by the \textbf{limit state
  function} $g(\vd,\mX)$, where $\vd\in\R{d_d}$ are the design variables, and
$\mX\in\R{d_r}$ are random variables.\cite{ditlevsen1996structural} For uniaxial
tension, we have the limit state function

\nomenclature[E]{$\R{}$}{Real Numbers}
\nomenclature[E]{$d$}{Dimension}
\nomenclature[D]{$g$}{Limit State}
\nomenclature[S]{$d$}{Design Variables}
\nomenclature[S]{$r$}{Random Variables}

\begin{equation}
  g(t,\mX) = U - F / A(t),
\end{equation}

\noindent where $g\leq0$ corresponds to failure, and $\mX = (U,F)^{\top}$
are the random variables, chosen to model different sources of uncertainty. The
critical ultimate stress $U$ replaces a fixed, deterministic stress
$\sigma_{ult}$ to model the variability inherent in manufacturing processes. The
applied load $F$ replaces a fixed load $f$ to model the uncertain conditions
the design will encounter. Reliable sizing is accomplished by solving the
optimization problem

\begin{equation} \begin{aligned} \label{eq:tension-exact}
    \text{min }  & C(t), \\
    \text{s.t. } & \P_{\mX}[g(t,\mX) > 0] \geq \cR,
\end{aligned} \end{equation}

\noindent where $C$ is the cost of the design, taken to be $C(t)=t$ for this
example, and $\cR\in[0,1]$ is the desired reliability. Here and below, we denote
by subscript the random variables considered in evaluating an expectation, e.g.
a probability statement. \Cref{eq:tension-exact} has an exact solution, defined
by

\nomenclature[D]{$C(\vd)$}{Design Cost}
\nomenclature[T]{$\cR$}{Reliability}
\nomenclature[O]{$\P_{\mX}[\cdot]$}{Probability; wrt $\mX$}
\nomenclature[T]{$\Phi(\cdot)$}{Standard Normal CDF}

\begin{equation} \begin{aligned} \label{eq:tension-exact-solution}
    A^* &= \frac{\mu_{u}\mu_f + \sqrt{\Phi^{-1}(\cR)^2\mu_{u}^2\tau_f^2%
        + \Phi^{-1}(\cR)^2\mu_f^2\tau_{u}^2%
        - \Phi^{-1}(\cR)^4\tau_{u}^2\tau_f^2}}%
           {\mu_{u}^2-\Phi^{-1}(\cR)^2\tau_{u}^2}, \\
    t^* &= \sqrt{A^*/\pi+r^2}-r, \\
\end{aligned} \end{equation}

\noindent where $\Phi^{-1}(\cdot)$ is the standard inverse normal CDF. In the
case where $r=1m$ and $\cR=0.95$, we find the solution $t^*\approx3.3cm$.

\subsection{Uncertain Parameters}
In practice, the parameters $\v\theta$ for the distribution of the random
variables may not be known. In the uniaxial tension example, we assume we know the
parameters for $F$ exactly, and have access to some number $m$ of samples
$\mU_i\sim \dN(\mu_{u},\tau^2_{u})$, which lead to the sample estimates
and their (sampling) distributions

\nomenclature[T]{$\v\phi$}{Distribution Parameters}
\nomenclature[E]{$m$}{Sample Count}
\nomenclature[O]{$\overline{\cdot}$}{Sample Mean}

\begin{equation}\begin{aligned}
  \overline{U} &= \frac{1}{m} \sum_{i=1}^m \mU_{i} %
                           \sim \dN(\mu_{u},\tau^2_{u}/m), \\
  S^2_{u} &= \frac{1}{m-1} \sum_{i=1}^m (\mU_{i} - \overline{U})^2 %
                           \sim \chi^2_{m-1}\tau^2_{u}/(m-1).
\end{aligned}\end{equation}

\noindent Note that we assume no additional measurement noise on the material
measurements $\mU_{i}$. Variation here is assumed to arise from
manufacturing variability alone. The parameter estimates
$\hat{\v\theta}=(\overline{U},S^2_{u})^{\top}$ are random and have the moments

\begin{equation} \begin{aligned}
    \E[\v\theta] &= (\mu_{u},\tau^2_{u})^{\top}, \\
    \Cov[\v\theta] &= \text{Diag}[\tau^2_{u}/m,\tau^4_{u}/(m-1)] \equiv \mT.
\end{aligned} \end{equation}

\noindent We will denote by $\hat{\mT}$ the sample estimate of $\Cov[\v\theta]$,
and will occasionally use a subscripted version $\hat{\mT}_m$ to emphasize the
sample size. The lack of perfect knowledge implies that exactly solving the RBDO
problem \eqref{eq:tension-exact} is not possible. Instead, one must turn to some
form of statistical approximation -- two possible approaches are detailed below.

\nomenclature[O]{$\hat{\cdot}$}{Sample Estimate}
\nomenclature[O]{$\text{Diag}(\cdot)$}{Diagonal Matrix}
\nomenclature[O]{$\E_{\mX}[\cdot]$}{Expectation; wrt $\mX$}
\nomenclature[O]{$\V_{\mX}[\cdot]$}{Variance; wrt $\mX$}
\nomenclature[O]{$\Cov_{\mX}[\cdot]$}{Covariance; wrt $\mX$}
\nomenclature[T]{$\mT$}{Covariance Matrix}

\subsection{Regulated and Mixed Design}
Under Title 14 CFR 25.613, commercial aircraft designers are required to
establish material properties using a \textbf{basis value}, a random variable
constructed from a random material population. Formally, a basis value is a
\textbf{tolerance interval}, a random interval constructed with respect to
another random variable, such that the interval contains a fraction $\cP$ of the
population at a desired confidence level $\cC$.\cite{meeker2017statistical} A
basis value is a one-sided interval, thus it is reported as a single number.

\nomenclature[T]{$\cP$}{Population Fraction}
\nomenclature[T]{$\cC$}{Confidence Level}
\nomenclature[T]{$B$}{Basis Value}

Practically, one may draw a number $m$ of samples of the desired material
property $U_{i}\sim\rho$ for $i=1,\dots,m$ and compute the sample mean
$\overline{U}$ and variance $S^2$. Effectively, the basis value is the mean
estimate, knocked down by the sample standard deviation, scaled by an
appropriate factor $k_{\cP,\cC}(m)$. Formally, we have

\begin{equation}
  B = \overline{U} - k_{\cP,\cC}(m) S,
\end{equation}

\noindent where $k_{\cP,\cC}(m)$ is determined by the desired Population
fraction $\cP$, Confidence level $\cC$, and chosen sample count $m$. Under a
normal $X$ assumption, the factor $k_{\cP,\cC}(N)$ can be determined from a
non-central t-distribution -- this assumption is exact in the uniaxial tension
problem defined above. One may also employ empirical methods for computing basis
values in the case of large sample sizes.\cite{meeker2017statistical}

Note that $B$ is a random variable, for which we compute a \emph{realization}
based on sample estimates. The basis value is applied by introducing a modified
limit state function

\begin{equation} \label{eq:tension-limit-bv}
    g(t,B,F) = B - F/A(t).
\end{equation}

\noindent Note that \eqref{eq:tension-limit-bv} is \emph{not} the true limit
state, but is instead an approximation induced by the basis value. We shall see
(Fig. \ref{fig:Meff-comp-bv-pi}) that this approximation \emph{will not
  necessarily lead to a conservative design}.

An additional level of conservatism is required by Title 14 CFR 25.303, which
imposes a factor of safety (FOS) of $1.5$ on external load limits. In this
\emph{regulated} approach to design, one sizes the cross-section via

\begin{equation} \label{eq:tension-regulated}
  A^*_{\text{regulated}} = B / (1.5 \mu_f),
\end{equation}

\noindent here using $\mu_f$ as the nominal loading conditions. We also pursue a
`mixed' approach using a basis value, which is (to our knowledge) not used in
industry, but better isolates the effect of the basis value approximation,
purely for illustrative purposes. Since the basis value is the only number
reported, we do not have enough information to evaluate the probability related
to the material population variability in \eqref{eq:tension-exact}. We instead
solve a modified optimization problem, given by

\begin{equation} \begin{aligned} \label{eq:tension-bv}
    \text{min }  & C(t), \\
    \text{s.t. } & R(B)\equiv\P_{F}[g(t,B,F) > 0] \geq \cR.
\end{aligned} \end{equation}

\noindent Note that the evaluated reliability $R(B)$ is now a random variable,
induced by the random basis value. Thus the $t$ which solves
\eqref{eq:tension-bv} is a random variable. Furthermore, the uncertainty arising
from the material property is \emph{not} accounted in the probability statement,
as implied by the subscript.

It is important to note that \emph{it is patently unreasonable to expect these
  approaches to compare favorably with true RBDO approaches} -- the regulated
and mixed approaches are simply not tailored for controlling failure
probabilities. However, we include these results to show how the regulated
approach fares in terms of realized system reliability. To our knowledge, such a
comparison has not been made in the literature -- the studies here give a sense
of what potential improvements could be made, should RBDO be more widely adopted
in real aircraft design. Intuitively, this potential for improvement exists
because, in both the regulated \eqref{eq:tension-regulated} and mixed
\eqref{eq:tension-bv} approaches, the material uncertainty is effectively
\emph{decoupled} from system reliability by the basis value.

\subsection{Plug-In Estimate}
As an alternative to the approaches above, one may model random material
properties,\cite{mil-hdbk-17-3e1997} estimate the distribution parameters
$\v\theta$, and evaluate all probabilities using the `plug-in' estimate
$\hat{\v\theta}$. This approach leads to the modified optimization problem

\begin{equation} \begin{aligned} \label{eq:tension-pi}
    \text{min }  & C(t), \\
    \text{s.t. } & R(\hat{\v\theta})\equiv%
                   \P_{\mX(\hat{\v\theta})}[g(t,\mX(\hat{\v\theta})) > 0] \geq \cR,
\end{aligned} \end{equation}

\noindent where we introduce the notation
$\mX(\hat{\v\theta})\sim\rho(\hat{\v\theta})$ to denote a random variable drawn
conditional on the assumed parameter values $\hat{\v\theta}$, and note that the
notation $\mX$ implies the random variable is drawn according to the ground
truth parameters $\v\theta$. Note that \eqref{eq:tension-pi} also involves a
random estimated reliability $R(\hat{\v\theta})$, with the randomness induced by
the estimated parameter values. Thus the $t$ which solves \eqref{eq:tension-pi}
is also a random variable. This design is computed using
\eqref{eq:tension-exact-solution}, substituting the estimated parameter values.

\subsection{Metrics and Results}
Here we compare the approaches above in terms of their performance, relative to
the exact solution of \eqref{eq:tension-exact}. For comparison, we introduce two
performance metrics; the \textbf{effective margin} $M_{\text{eff},\cC}(\vd)$ and
\textbf{effective reliability} $R_{\text{eff},g}(\vd)$, defined in
\eqref{eq:effective-metrics} below.

\begin{equation} \begin{aligned} \label{eq:effective-metrics}
    M_{\text{eff},C}(\vd) &\equiv \frac{C(\vd)-C^*}{C^*}, \\
    R_{\text{eff},g}(\vd) &\equiv \P_{\mX}[g(\vd,\mX) > 0].
\end{aligned} \end{equation}

\noindent Note that since $M_{\text{eff},C}(\vd)$ is defined with respect to a
minimal objective value $C^*$, it is only defined for RBDO problems where such a
value exists. The effective margin $M_{\text{eff},C}(\vd)$ measures
\emph{system} performance in terms of the chosen cost metric $C(\vd)$. If
$M_{\text{eff},C}(\vd)$ is positive, it implies there must be slackness in the
reliability constraints (under the true parameter values $\v\theta$), and the
cost of the design could be reduced. Conversely, negative effective margin
implies the cost observed could not have been achieved without violating a
constraint -- in this case effective margin is an indication of how
\emph{under}-built a design is.

The effective reliability $R_{\text{eff},g}(\vd)$ directly measures the achieved
reliability of a design, in terms of a \emph{single constraint}. An
$R_{\text{eff},g}(\vd)$ less than (greater than) the desired reliability implies
under- (over-) design in the system. In contrast with effective margin, which
gives a single measure for a design problem, one would have a set of effective
reliabilities for a design problem -- one for each reliability constraint. We
will illustrate a case with multiple constraints below.

Note that we will use these quantities to measure the performance of
\emph{design strategies} by considering an ensemble of random
designs\footnote{We have found that some have difficulty accepting the concept
  of \emph{random designs}. Note that any deterministic function or process,
  given a random input, \emph{necessarily} produces a random output.} arising
from different approaches. Furthermore, these quantities are \emph{frequentist}
constructions, as they are predicated on the existence of a true parameter value
$\v\theta$.

\begin{figure}[!ht]
  \centering
  \includegraphics[width=0.65\textwidth]{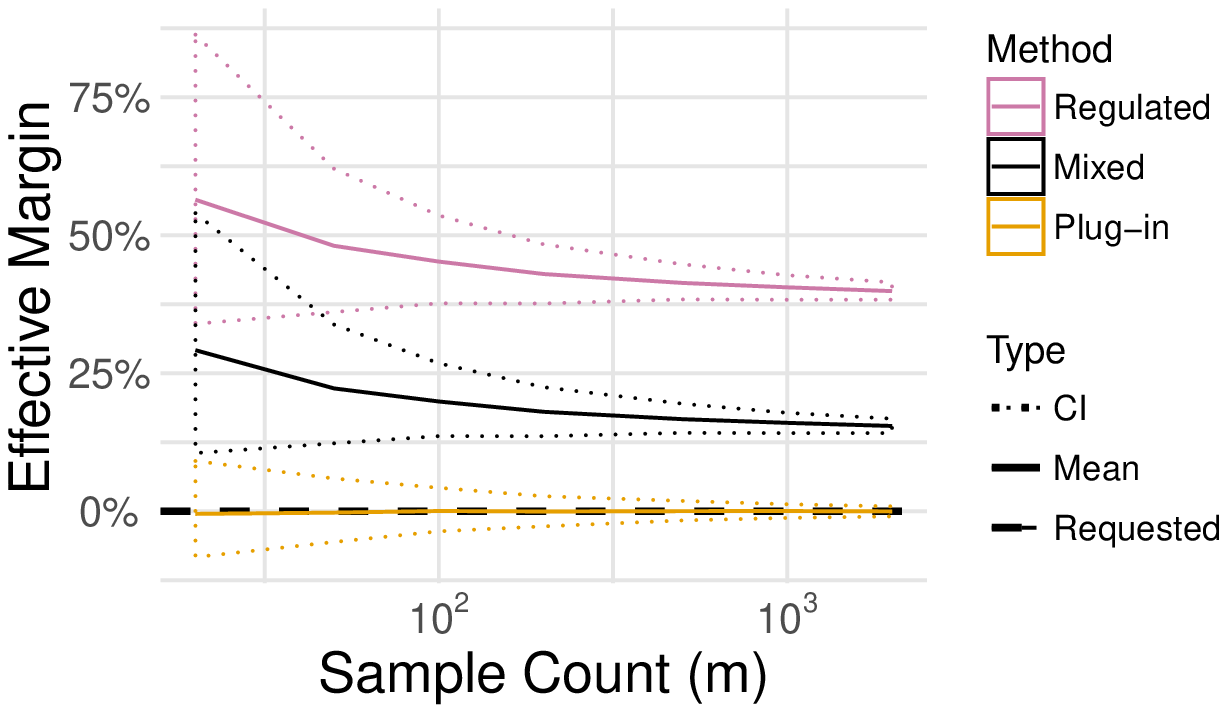}
  \includegraphics[width=0.65\textwidth]{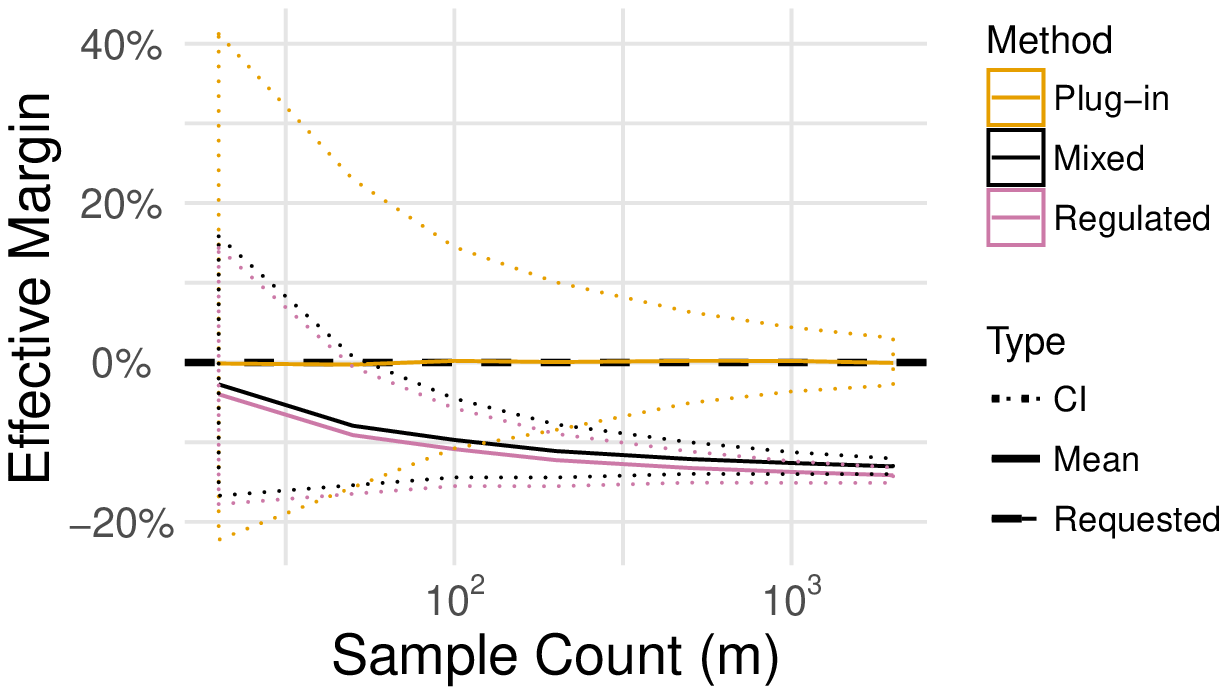}

  \caption{Effective margin against sample size for designed reliability
    $\cR=0.99$ (Top) and $\cR=1-10^{-7}$ (Bottom) in the uniaxial tension
    example. Since the reliable design problems are solved analytically, all
    pathologies arise from the materials characterization process. We use an
    A-basis value in both reliability problems. All approaches necessarily
    return random designs due to material uncertainties: Mean profiles and
    two-sided $95\%$ confidence intervals are approximated using $10^3$
    replications. The results shown here illustrate that the regulated and mixed
    approaches do not control the failure probability. At low reliability (Top),
    the use of a basis value leads to \emph{unintentional} effective margin,
    while at high reliability (Bottom) its use prevents the desired reliability
    from being achieved. Since no margin is added to the design, any effective
    margin (whether positive or negative) is unintentional, and opaque to the
    designer. The plug-in approach has zero effective margin in the asymptotic
    limit, but returns an unacceptable fraction of under-performing designs at
    reasonable sample sizes. Ideally, one would desire a design procedure which
    has positive effective margin at some designed confidence level. In Sec.
    \ref{sec:sampling-margin} we introduce a procedure which approaches the
    ideal.}
  \label{fig:Meff-comp-bv-pi}
\end{figure}

Since the $t$ arising from the strategies above are random, the resulting
performance metrics are also random. We simulate the sizing problem by drawing a
variable number of material samples $m$, solving the optimization problems
analytically, and replicate this entire procedure to build confidence intervals
that measure \emph{design strategy} performance. The results shown in Figure
\ref{fig:Meff-comp-bv-pi} demonstrate deficient behavior with all approaches
discussed above.

Both the regulated and mixed approaches result in either over- or under-designed
solutions, depending on the desired reliability. Intuitively, this deficiency is
due to `decoupling' of attendant uncertainties from the system reliability. In
computing a basis value, one gathers enough data to estimate the mean and
variance of a material population, but then collapses all data to a single
number for structural sizing. Any following design for reliability cannot
account for distributional information in this framework, which results in a
lack of control over the ultimate failure chance. Stated differently, the basis
value approach attempts to add a form of margin (in the $-k_{\cP,\cC}(m)S$ term)
to the material property, and additional forms of margin are added in the
downstream design process; since these margins are not designed in terms of the
system reliability, it is unsurprising they fail to control the system failure
chance.

Note that given two standards of basis value -- A- and B-basis -- and the
modeling assumptions used to generate them (random variable model and sample
size), one can easly recover the estimated moments of the data, and use these
for reliability design. However, one cannot reasonably claim to be performing
design \emph{using} basis values in this case, as the results will be identical
to the plug-in approach.

The plug-in approach asymptotically recovers zero effective margin, but returns
an unacceptable fraction of negative effective margin designs. This is because
the plug-in approach adds \emph{no} form of margin. The estimated parameter
values are assumed to be true for the purposes of sizing; when the material
capacity mean is overestimated (or the variance underestimated), the resulting
design will be less reliable than desired. In practice, a designer would want a
\emph{principled} way to add margin to quantities directly related to failure
criteria. These results motivate the introduction of \textbf{\margin{}}.

\begin{figure}[!ht]
  \centering
  \includegraphics[width=0.65\textwidth]{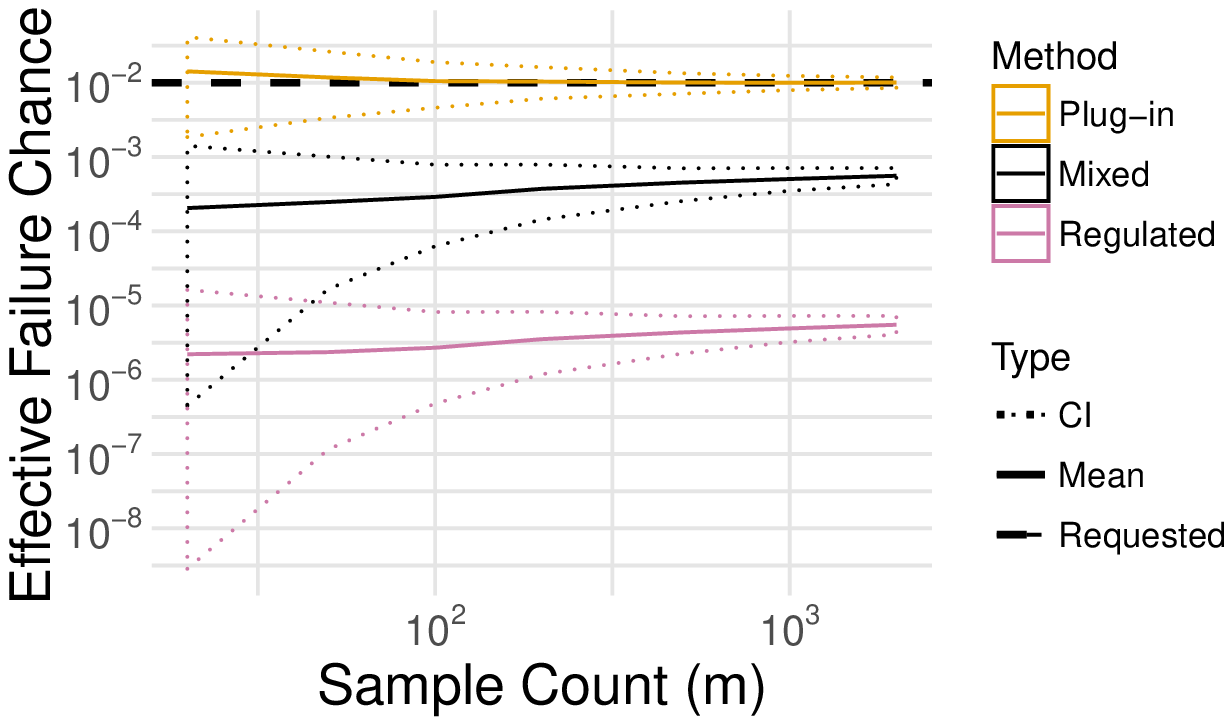} \\
  \includegraphics[width=0.65\textwidth]{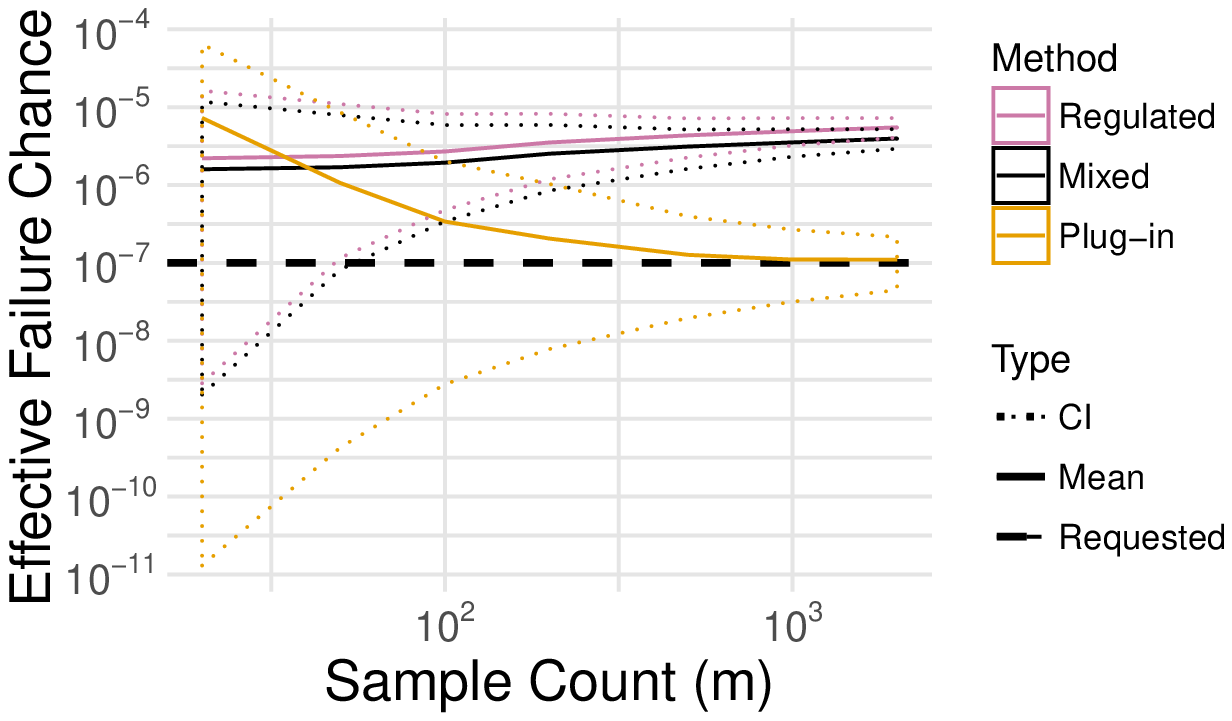}

  \caption{Effective reliability against sample size for designed reliability
    $\cR=0.99$ (Top) and $\cR=1-10^{-7}$ (Bottom) in the uniaxial tension
    example. For ease of plotting, we report the \textbf{effective failure
      chance} $F_{\text{eff}}=1-R_{\text{eff}}$, which carries the same
    information. The conclusions of Figure \ref{fig:Meff-comp-bv-pi} are echoed
    here. At low reliability targets the use of a basis value results in an
    overly-conservative design; that is, the failure chance is lower than
    requested, implying that material could be removed and the design would
    still achieve the desired reliability. Despite the fact that the A-basis
    value population fraction matches the desired reliability target, both
    approaches using a basis value lead to significant over-design. At high
    reliability targets, the opposite issue arises; the failure chance is higher
    by orders-of-magnitude.}
  \label{fig:Feff-comp-bv-pi}
\end{figure}

\FloatBarrier
\section{\MARGIN{}} \label{sec:sampling-margin}
In this section, we present a design methodology which overcomes the issues
inherent in both the basis value and plug-in approaches. Here we introduce the
general concept of \textbf{\margin{}}, provide examples of its implementation, and
present results for the uniaxial tension sizing problem.

\subsection{\MARGIN{} Concept}
\textbf{Margin} is a simple but ubiquitous concept from engineering. Margin is a
displaced threshold for some constraint, added to encourage a conservative
design. Within the RBDO framework, one can add margin in at least two ways:

\begin{equation} \begin{aligned} \label{eq:exact-margin}
    \P_{\mX}[g(\vd,\mX) > g_m] &\geq \cR, \\
    \P_{\mX}[g(\vd,\mX) > 0] &\geq \cR+p,
\end{aligned} \end{equation}

\noindent which we refer to (respectively) as margin in limit (MIL) state, and
margin in probability (MIP): We will see below that the MIP formulation provides
additional, desirable properties. In \eqref{eq:exact-margin}, adding positive
margin ($g_m,p>0$) will result in an overly-conservative design, with regard to
the desired reliability. However, adding margin is useful in the realistic case
where the parameters $\v\theta$ are not exactly known.

We introduce the concept of \textbf{\margin{}} as a form of margin added to handle
the statistical uncertainties in $\hat{\v\theta}$ arising from an estimation
procedure. Thus, we introduce the following definition:

\bigskip
\noindent\underline{Definition:} \Margin{} (\TLA{}) is any form of margin which:

\begin{enumerate}
  \item Improves the reliability of a system limit state, based on
    discrepancy with the realized reliability \label{def:sm1}
  \item Decays to zero with increased precision \label{def:sm2}
\end{enumerate}

\bigskip
These requirements are inspired both by the deficiencies found among the methods
in Section \ref{sec:motivation} and by existing approaches in
literature.\cite{ito2018conservative} The plug-in approach uses estimates for
`best guess' parameter values, but does not account for how those estimates may
affect the realized reliability. Conversely, both basis value approaches add
some form of margin, but with a value decoupled from the system reliability.
Point \ref{def:sm1} addresses these deficiencies. Note that the regulated and
mixed approaches also failed to converge to the true system reliability, even as
the number of samples $m$ approached infinity. Point \ref{def:sm2} addresses
this, by imposing a convergence criteria.

Note that \margin{} is intended to deal with statistical uncertainties
\emph{only}; this excludes unidentified uncertainties. This flexible definition
can be implemented in multiple ways, as illustrated below.

\subsection{Margin in Limit}
Here we define the margin in limit (MIL) as a margin term $g_{MIL,\cC}$ based on
the mean difference between the estimated limit state $\hat{g}$ and the true
limit state $g$. This margin is defined at a desired confidence level $\cC$ by

\begin{equation} \label{eq:MIL-def}
  \P_{\hat{\v\theta}}\left[g_{MIL,\cC} > E_{\mX(\hat{\v\theta})}[g(\vd,\mX(\hat{\v\theta}))] -
    \E_{\mX}[g(\vd,\mX(\v\theta))]\right] = \cC.
\end{equation}

\noindent Note that $E_{\mX(\hat{\v\theta})}[g(\vd,\mX(\hat{\v\theta}))]$ is a
random variable, due to the randomness induced by $\hat{\v\theta}$. In order for
$g_{MIL,\cC}$ to be a \TLA{}, it must converge to zero as
$\hat{\v\theta}\to\v\theta$. We present a proof of this fact in Appendix
\ref{sec:MIL-pf}. For the uniaxial tension example, the margin in limit \TLA{}
has an analytic expression, given by $g_{MIL,\cC} =
\Phi^{-1}(\cC)\tau_{u}/\sqrt{m}$, independent of the thickness $t$. In a more
general setting $g_{MIL,\cC}$ may be a function of the design variables $\vd$, a
fact which has implications for RBDO, and which will be revisited in in Section
\ref{sec:estimation}.

\begin{figure}[!ht]
  \centering
  \includegraphics[width=0.65\textwidth]{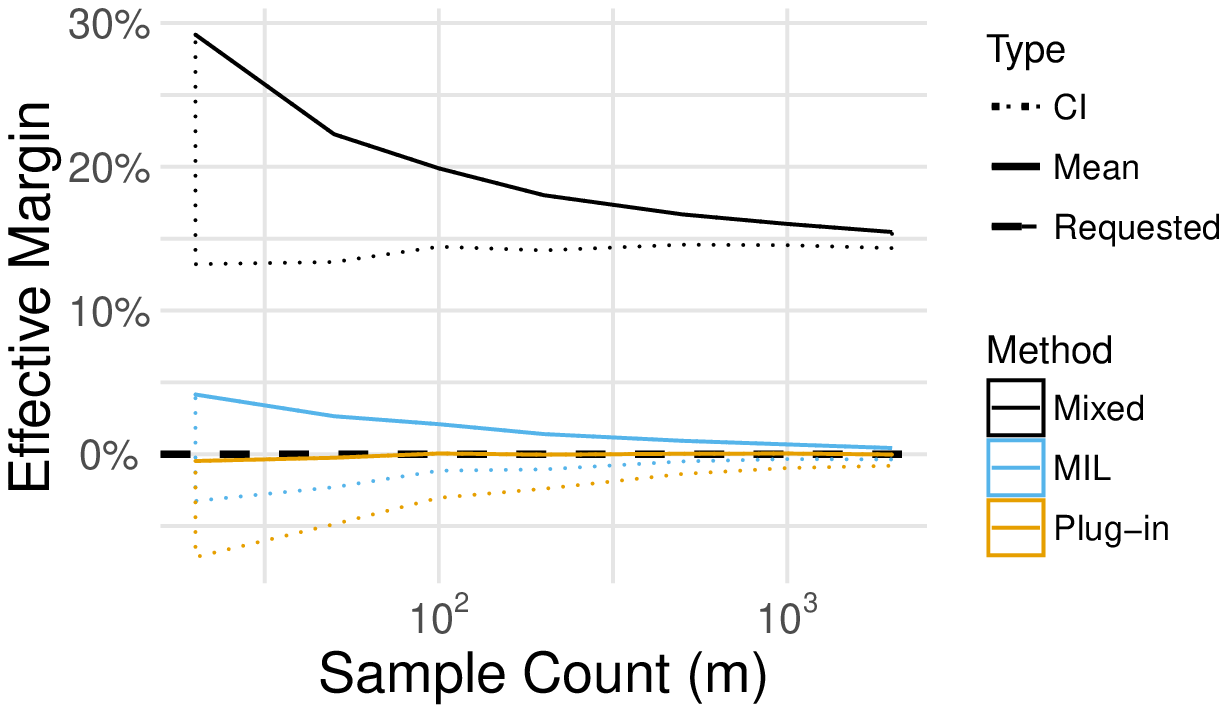} \\
  \includegraphics[width=0.65\textwidth]{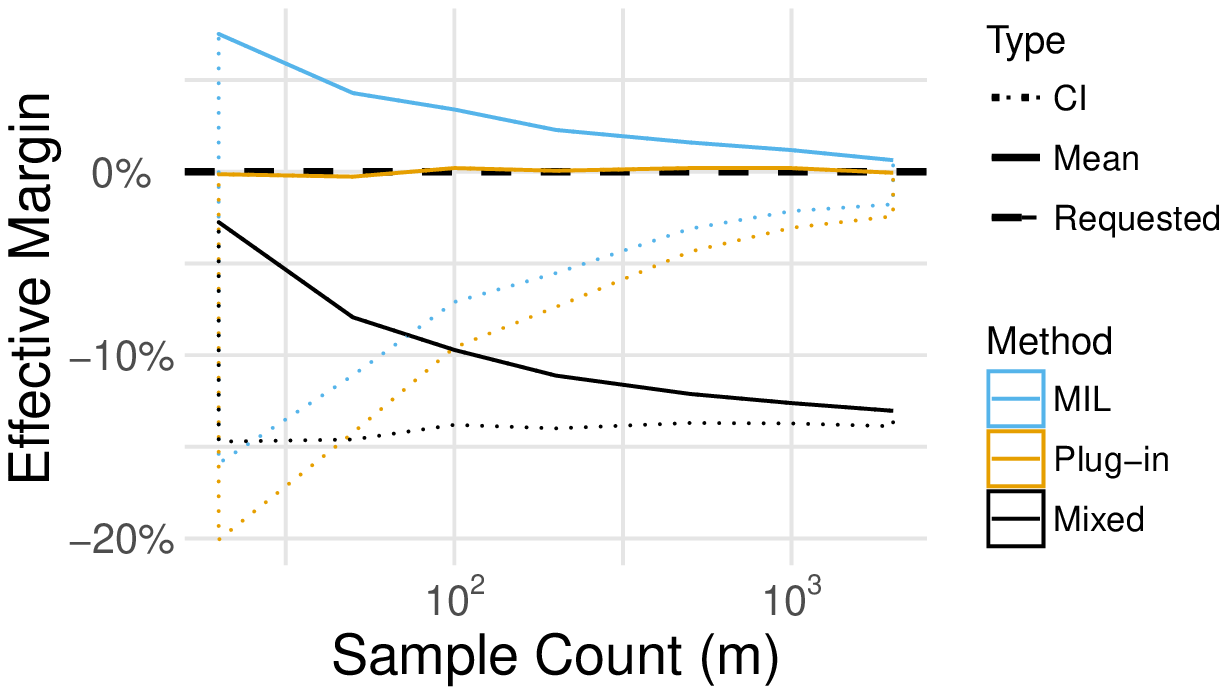}

  \caption{Comparison of margin in limit \TLA{} (MIL \TLA{}) against an A-Basis
    Value (BV) and Plug-In (PI) approaches at $\cR=0.99$ (Top) and
    $\cR=1-10^{-7}$ (Bottom). Probabilities are evaluated analytically at
    estimated parameter values, and $10^3$ replications are carried out to
    construct mean curves and one-sided $95\%$ confidence intervals. As
    predicted, the MIL approach is more conservative than the plug-in approach.
    Crucially, the margin in limit approach approaches the desired reliability
    as the sample count is increased, in contrast with the regulated and mixed
    approaches. Note that the MIL approach demonstrated here relies on exact
    knowledge of $\v\theta$; we present an approximation of this approach in
    Sec. \ref{sec:estimation}. Despite the use of exact knowledge, even the MIL
    approach leads to an unacceptable fraction of under-performing designs,
    particularly in the strict-reliability case; in Sec. \ref{sec:estimation} we
    introduce an alternative strategy which addresses this issue.}
  \label{fig:MIL-tension}
\end{figure}

Example results shown in Figure \ref{fig:MIL-tension} demonstrate that the mean
difference \TLA{} is indeed more conservative than the plug-in approach, but
does not guarantee non-zero effective margin at the desired confidence level, a
property we will achieve with a different implementation below. Crucially, the
margin in limit \TLA{} results approach the desired reliability, in contrast
with the approaches employing basis values. Note also that the margin in limit
\TLA{} formally relies on exact knowledge of $\v\theta$; we will introduce an
approximation to this margin term below.

\subsection{Margin in Probability}
An equally valid means to add margin is to apply margin in the estimated
reliability, as in the second line of \eqref{eq:exact-margin}. This is an
attractive option, as it more directly controls the quantity of interest for
design for reliability -- the failure chance -- rather than exerting an indirect
influence through the limit state. Margin in probability $p$ is applied by
designing for the modified constraint

\begin{equation} \label{eq:pr-constraint}
  R(\hat{\v\theta}) = \P_{\mX(\hat{\v\theta})}[g(\vd,\mX(\hat{\v\theta})) > 0] \geq \cR + p,
\end{equation}

\noindent where $p$ is determined via the coupled auxiliary equation

\begin{equation}
  \P_{\hat{\v\theta}}[p > R(\hat{\v\theta}) - R(\v\theta)] = \cC.
\end{equation}

\noindent Applying margin in this fashion has a very desirable property; in this
form, the confidence level $\cC$ can be interpreted as a probability of
satisfying the desired reliability $R(\v\theta,\vd(\hat{\v\theta}))\geq\cR$
\emph{over the distribution of random designs}. We can see this by first
assuming a slack form of the reliability constraint \eqref{eq:pr-constraint} is
satisfied,

\begin{equation}
  R(\hat{\vd},\hat{\v\theta}) = \cR + p + \epsilon,
\end{equation}

for any given random design $\hat{\vd}\equiv\vd(\hat{\v\theta})$, and computing

\begin{equation} \begin{aligned} \label{eq:R-coverage}
    \cC &= \P_{\hat{\v\theta}}[p > R(\hat{\v\theta},\hat{\vd}) - R(\v\theta,\hat{\vd})], \\
        &= \P_{\hat{\v\theta}}[R(\v\theta,\hat{\vd}) > \cR + \epsilon].
\end{aligned} \end{equation}

\noindent Interpreting the probabilities of \eqref{eq:R-coverage} requires that
we consider random designs arising from the employed \emph{design strategy}.
Since the design $\hat{\vd}$ is random (induced by the random parameters
$\hat{\v\theta}$), this allows us to interpret the probability over
$\hat{\v\theta}$.

We use the term \textbf{confidently conservative} (C2) to denote a design
\emph{strategy} with the property $\cC =
\P_{\hat{\v\theta}}[R(\v\theta,\hat{\vd}) > \cR]$. A strategy which is C2 is
conservative in reliability at a known confidence level $\cC$. In the case where
the reliability constraint is not slack (i.e. $\epsilon=0$) over the
distribution of $\hat{\vd}$, the MIP strategy is C2. For the MIP strategy, and
by the non-decreasing property of CDF's, a slack reliability constraint implies
a higher confidence level, while an infeasible constraint implies a lower
confidence level.

Note that while C2 is a desirable property, we do \emph{not} demand that a
\TLA{} be C2; this is because the property will be practically unattainable in
any real engineering context, due to challenges such as unknown unknowns. We
introduce C2 as a theoretical ideal that practical design strategies can
approach. The example below will illustrate the C2 property of this design
strategy.

Here we draw a comparison with the \textbf{conservative reliability index} (CRI)
of Ito et al.,\cite{ito2018conservative} which is closely related to our
proposed MIP approach. In nomenclature consistent with our presentation, they
recommend solving

\begin{equation} \begin{aligned} \label{eq:cri}
    \text{min. } &C(\vd), \\
    \text{s.t. } &R^{\alpha} \geq \cR, \\
                 &\P_{\hat{\theta}}[R(\hat{\v\theta}) > R^{\alpha}] = \alpha. \\
\end{aligned} \end{equation}

\noindent We note that the CRI approach is a form of \margin{}, as it encourages
conservatism based on the variability in the estimated reliability, and indeed
recovers the true reliability with perfect information (Appendix
\ref{sec:CRI-pf}). However, with this formulation, we arrive not at a C2
condition, but rather find that

\begin{equation} \label{eq:cri-outcome}
  \P_{\hat{\v\theta}}[R(\hat{\v\theta}) > \cR] = \alpha,
\end{equation}

\noindent which is the reliability conditional on the \emph{estimated} parameter
values which, for $\v\theta$ which take continuous values, will be correct with
zero probability. One implication of the CRI approach is that bias in the
estimated parameters can cause significant depatures in the realized reliability
$R(\v\theta)$ from the desired threshold $\cR$; we illustrate this fact with a
simple example in Appendix \ref{sec:ex-bias}.

Despite directly controlling the reliability, applying margin in probability has
a weakness -- this approach is more numerically unstable than applying margin
directly to the limit state. If $p$ is estimated via some noisy procedure, then
it is possible for $\cR+p\geq1$ to occur. In this case, the resulting
reliability problem is ill posed. The example below will also illustrate this
pathology.

\begin{figure}[!ht]
  \centering
  \includegraphics[width=0.65\textwidth]{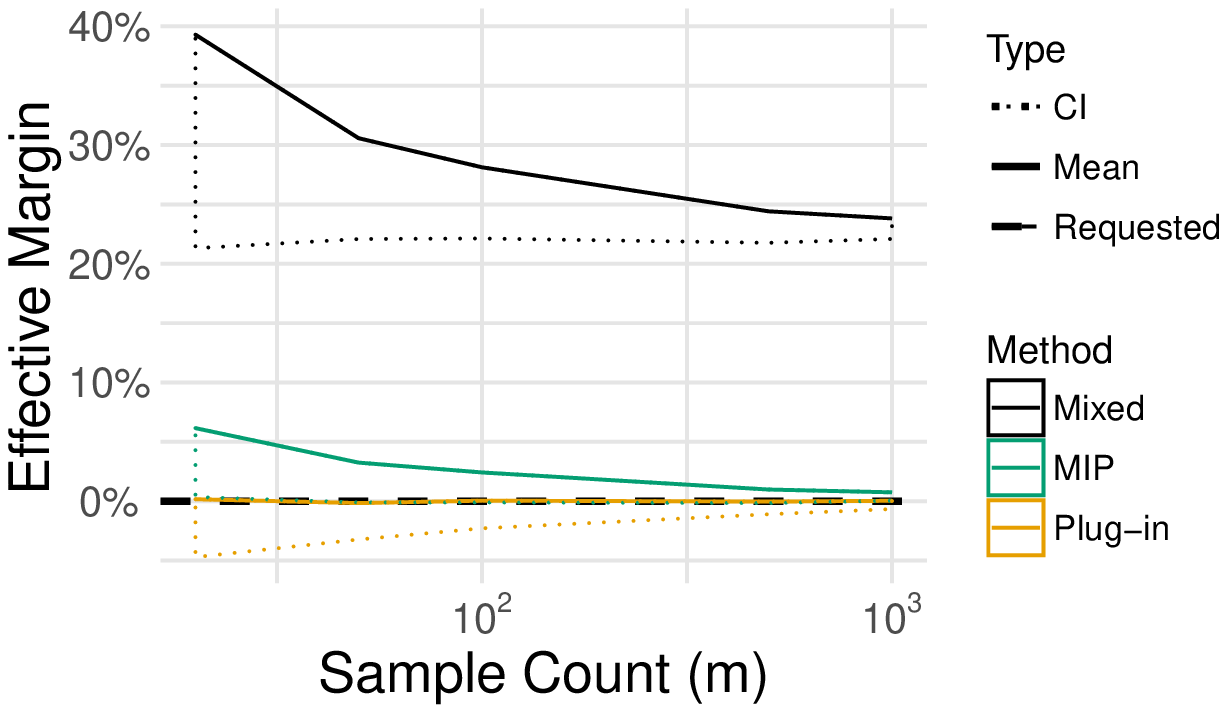} \\
  \includegraphics[width=0.65\textwidth]{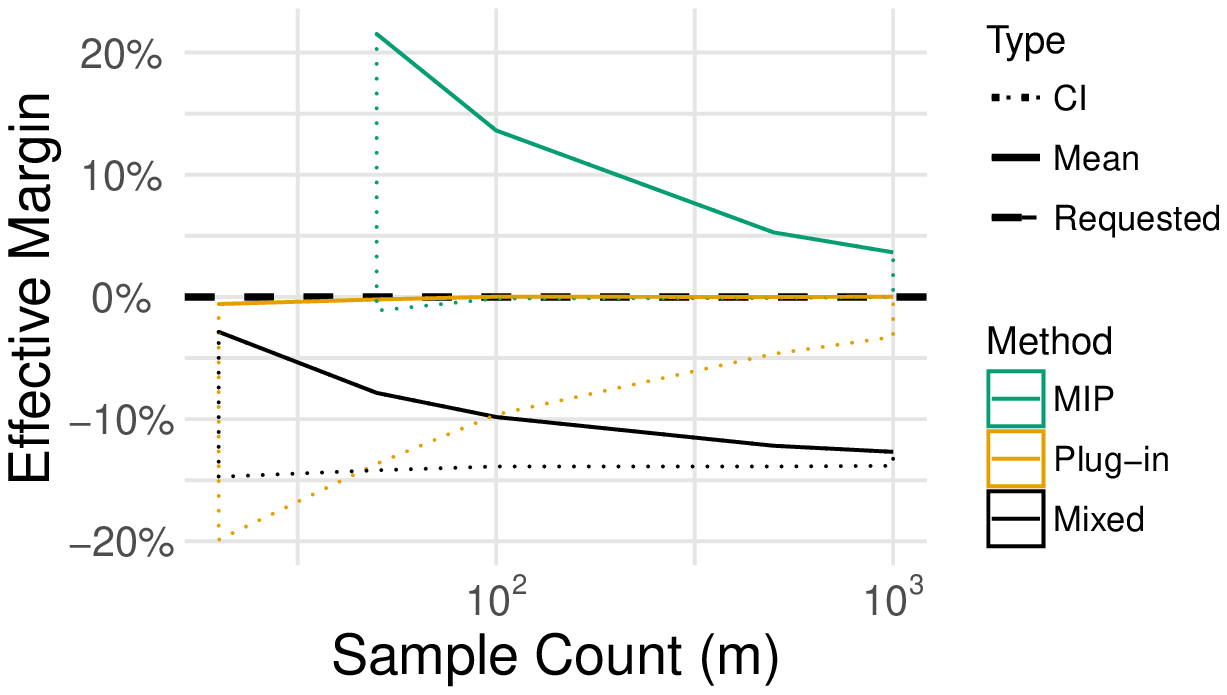}

  \caption{Comparison of margin in probability (MIP) against an A-Basis Value
    and Plug-In approaches at $\cR=0.99$ (Top) and $\cR=1-10^{-7}$ (Bottom). The
    MIP approach is carried out semi-analytically. The MIP results are similar
    to those of the MIL \TLA{} (Fig. \ref{fig:MIL-tension}), but demonstrate the
    confidently conservative property defined above. Note that at lower sample
    counts ($m<50$) in the strict reliability case (Bottom), results are not
    given for low sample counts. This is due to realizations where the estimated
    margin is incompatible with the desired reliability (i.e. $\cR+p>1$) -- this
    strategy is C2 contingent on the constraint $R(\hat{\v\theta})\geq\cR + p$.
    This pathology illustrates a point: Margin in probability is a more
    numerically unstable procedure, as compared with other forms of margin
    suggested in this work. An alternative (positive) view on the phenomenon is
    that margin in probability can signal that the available information is
    incompatible with the desired reliability targets. In practice, a designer
    may use MIP to determine when additional precision in estimates is
    required.}
  \label{fig:MIP-tension}
\end{figure}

In the tension sizing example, the estimated reliability has an analytical form

\begin{equation}
  R(\hat{\v\theta}) = \Phi\left(\frac{\overline{X}-\mu_f/A(d)}%
                                     {\sqrt{S^2+\tau_f^2/A(d)}}\right),
\end{equation}

\noindent which we use in a semi-analytic study of the probability margin
approach, solving the design problem via fixed-point iteration. Results of this
numerical demonstration are reported in Figure \ref{fig:MIP-tension},
demonstrating the C2 property described above.

\FloatBarrier
\section{Enabling Estimation} \label{sec:estimation}
The implementations of \TLA{} above are intractable for realistic problems, as
they rely on knowledge of the unknown parameters $\v\theta$, and utilize exact
reliability evaluations or expensive second-order Monte Carlo approximations.
This section builds up the tools necessary to enable estimation of the two
\TLA{} implementations introduced above, using only information available
through the estimated parameters $\hat{\v\theta}$ and limit state function
evaluations $g(\vd,\mX_i(\hat{\v\theta}))$. The key insight is to build a random
variable model of our margin terms, justified by the delta method and enabled by
an efficient gradient approximation technique.

\subsection{Delta Method} \label{sec:delta-method}
The delta method is a classical result from the statistics community, and is
frequently used to estimate moments and construct confidence
intervals.\cite{van1998asymptotic} A theorem sufficient for our purposes is
stated here.

\bigskip
\noindent\underline{Theorem:} Let $\phi:\R{d_p}\to\R{}$ be differentiable at
$\v\theta\in\R{d_p}$, and let $\hat{\v\theta}\sim \dN(\v\theta,\mT_m)$ be a random
vector with $\mT_m\to0$ as $m\to\infty$. Then
$\phi(\hat{\v\theta})\stackrel{d}{\to}
\dN(\v\theta,\left.\nabla_{\v\theta}\phi\right|_{\v\theta}^{\top}
\mT_m\left.\nabla_{\v\theta}\phi\right|_{\v\theta})$ as $m\to\infty$, where
$\stackrel{d}{\to}$ denotes convergence in distribution.

\bigskip
The theorem above can be understood in terms of a first-order Taylor
approximation to the function $\phi(\hat{\v\theta})\approx\phi(\v\theta)+
\nabla_{\v\theta}\phi^{\top}(\hat{\v\theta}-\v\theta)$, which has a mean and
variance matrix matching the normal distribution above. As the estimator
$\hat{\v\theta}$ concentrates towards $\v\theta$ with increasing $m$ (implied by
its shrinking covariance matrix), the first-order approximation becomes more
accurate, providing an intuitive explanation of the delta method. Note that more
general results may be employed for non-normal cases, so long as a similar
convergence criterion is met.\cite{van1998asymptotic}

Crucially, the result above implies that, under the stated conditions, a
function of our estimated parameters $\hat{\v\theta}$ is asymptotically normal
-- an implication which we may use to build a model of our margin terms. We will
employ plug-in estimates for the parameters ($\v\theta,\mT$), which leaves the
gradient remaining to estimate.

\subsection{Parameter Gradients} \label{sec:gradient-trick}
A simple means to approximate the gradient would be a finite difference
approximation. However, this approach would be problematic if the mean
difference were approximated via Monte Carlo sampling. For example, if $n$
samples were employed to estimate $g_{MIL,\cC}$, an additional $n\times d_r$
samples would be required to approximate
$\left.\nabla_{\hat{\v\theta}}g_{MIL,\cC}\right|_{\hat{\v\theta}}$. Furthermore,
the computational noise arising from Monte Carlo estimation would necessitate a
careful choice of finite difference step size.\cite{more2012}

Rather than employ finite differences, we leverage the analytic form of the
modeled random variable $\rho(\hat{\v\theta})$ in the likelihood ratio (LR)
approach.\cite{l1990unified} Note that both the mean difference and probability
margins are defined in terms of an \emph{expectation}; we will first consider
the general case, and then specialize the results below.

Let

\begin{equation} \begin{aligned}
    \phi(\hat{\v\theta}) &= \E_{\mX(\hat{\v\theta})}[f(\mX(\hat{\v\theta}))], \\
      &= \int f(\mX)\rho(\mX;\hat{\v\theta})d\mX,
\end{aligned} \end{equation}

\noindent and note that $\phi(\hat{\v\theta})$ depends on its argument only
through the distribution PDF; that is, not through $f(\cdot)$ directly. Thus,
we may manipulate the gradient

\begin{equation} \begin{aligned} \label{eq:param-grad}
    \left.\nabla_{\hat{\v\theta}}\phi\right|_{\hat{\v\theta}} &= %
          \nabla_{\hat{\v\theta}}\int f(\mX)\rho(\mX;\hat{\v\theta})d\mX, \\
       &= \int f(\mX)\nabla_{\hat{\v\theta}}\rho(\mX;\hat{\v\theta})d\mX, \\
       &= \int f(\mX)\frac{\nabla_{\hat{\v\theta}}\rho(\mX;\hat{\v\theta})}%
                {\rho(\mX;\hat{\v\theta})}\rho(\mX;\hat{\v\theta})d\mX, \\
       &= \E_{\mX(\hat{\v\theta})}\left[f(\mX)%
          \frac{\nabla_{\hat{\v\theta}}\rho(\mX;\hat{\v\theta})}%
          {\rho(\mX;\hat{\v\theta})}\right],
\end{aligned} \end{equation}

\noindent which is an expectation with respect to the same density
$\rho(\hat{\v\theta})$, but with a modified integrand. The quantity
$\nabla_{\hat{\v\theta}}\rho(\mX;\hat{\v\theta})/\rho(\mX;\hat{\v\theta})$ is
known as the \textbf{score function}.\cite{van1998asymptotic} At first,
expectation \eqref{eq:param-grad} may appear to be a new quantity requiring a
separate Monte Carlo estimate, which would double the expense of approximating
$\phi(\hat{\v\theta})$ alone. However, note that $f(\mX)$ is unchanged within
the expectation of \eqref{eq:param-grad}; the parameter sensitivity is
represented by the score. If $\phi$ were approximated via Monte Carlo

\begin{equation}
  \phi(\hat{\v\theta}) \approx \frac{1}{m}\sum_{i=1}^m f(\mX_i(\hat{\v\theta})),
\end{equation}

\noindent with $\mX_i(\hat{\v\theta})\sim\rho(\hat{\v\theta})$, then we may
approximate the gradient \emph{using the same samples} via

\begin{equation}
  \left.\nabla_{\hat{\v\theta}}\phi\right|_{\hat{\v\theta}} \approx %
  \frac{1}{m}\sum_{i=1}^m f(\mX_i(\hat{\v\theta})) %
    \frac{\nabla_{\hat{\v\theta}}\rho(\mX_i(\hat{\v\theta});\hat{\v\theta})}%
       {\rho(\mX_i(\hat{\v\theta});\hat{\v\theta})}.
\end{equation}

\noindent Since the evaluation of $f$ is usually the limiting computation, this
procedure adds virtually no additional computational expense.

\subsection{Modeling the Margin in Limit} \label{sec:MIL-est}
The parameter gradient may be employed to model and estimate the margin in limit
in an economical fashion. Let

\begin{equation} \begin{aligned}
    D(\vd,\hat{\v\theta}) &= \E_{\mX(\hat{\v\theta})}[g(\vd,\mX(\hat{\v\theta}))]
                          - \E_{\mX}[g(\vd,\mX)], \\
                        &= \E_{\mX(\hat{\v\theta})}\left[
                           g(\vd,\mX(\hat{\v\theta})) - \E_{\mX}[g(\vd,\mX)]\right],
\end{aligned} \end{equation}

\noindent which has the parameter gradient

\begin{equation} \label{eq:MIL-gradient}
  \left.\nabla_{\hat{\v\theta}}D\right|_{\vd,\hat{\v\theta}} =
  \E_{\mX(\hat{\v\theta})}\left[\left(g(\vd,\mX(\hat{\v\theta})) - \E_{\mX}[g(\vd,\mX)]\right) %
    \frac{\nabla_{\hat{\v\theta}}\rho(\hat{\v\theta})}{\rho(\hat{\v\theta})}\right],
\end{equation}

\noindent which enables first-order approximation of the moments

\begin{equation} \begin{aligned} \label{eq:MIL-moments}
    \mu_D(\vd,\hat{\v\theta}) &\approx \tilde{\mu}_D(\vd,\hat{\v\theta}) = 0, \\
    \tau_D(\vd,\hat{\v\theta})^2 &\approx \tilde{\tau}_D(\vd,\hat{\v\theta})^2 \equiv
           \left.\nabla_{\hat{\v\theta}}D\right|_{\vd,\hat{\v\theta}}^{\top}\mT_m
           \left.\nabla_{\hat{\v\theta}}D\right|_{\vd,\hat{\v\theta}}.
\end{aligned} \end{equation}

\noindent As noted above, in the case where
$\hat{\v\theta}\stackrel{d}{\to}\dN(\v\theta,\mT_m)$, we find that
$D(\vd,\hat{\v\theta})$ is asymptotically normal. This justifies a model for the
margin in limit \TLA{}

\begin{equation} \label{eq:MIL-model}
  \P_{Z}[\tilde{g}_{MIL,\cC} > \tilde{\mu}_D(\vd,\hat{\v\theta})
                        + Z\tilde{\tau}_D(\vd,\hat{\v\theta})] = \cC,
\end{equation}

\noindent with $Z\sim \dN(0,1)$. This model problem has the exact solution

\begin{equation}
  \tilde{g}_{MIL,\cC} = \Phi^{-1}(\cC)\tilde{\tau}_D(\vd,\hat{\v\theta}).
\end{equation}

\begin{figure}[!ht]
  \centering
  \includegraphics[width=0.65\textwidth]{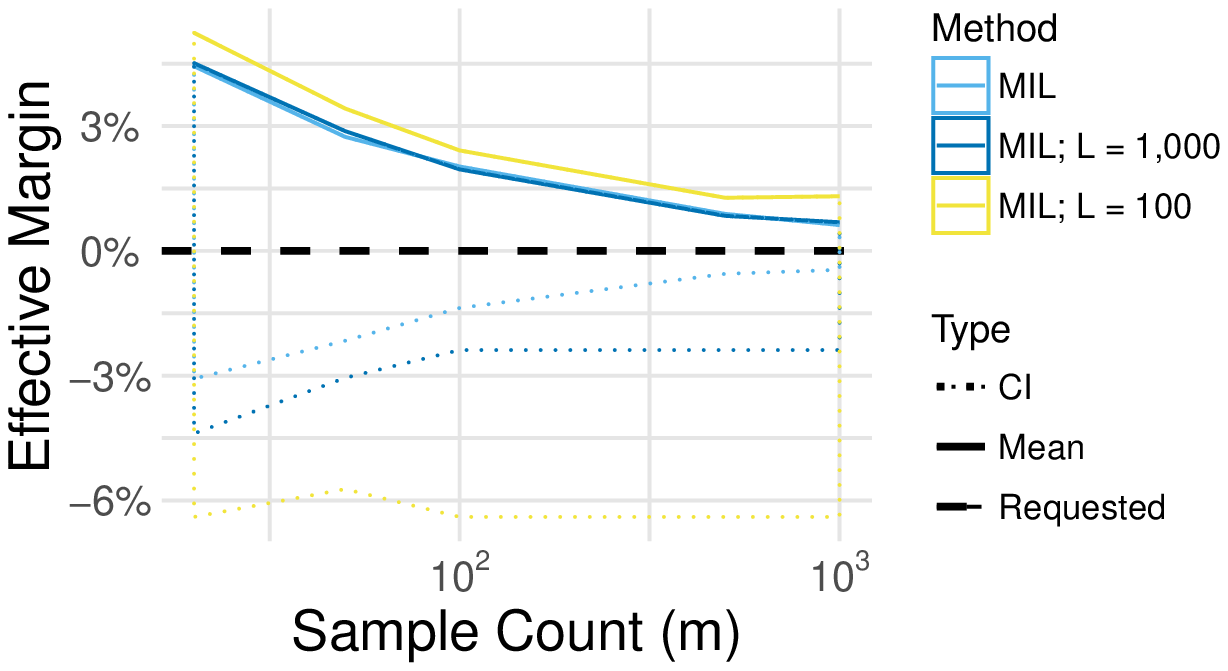} \\
  \includegraphics[width=0.65\textwidth]{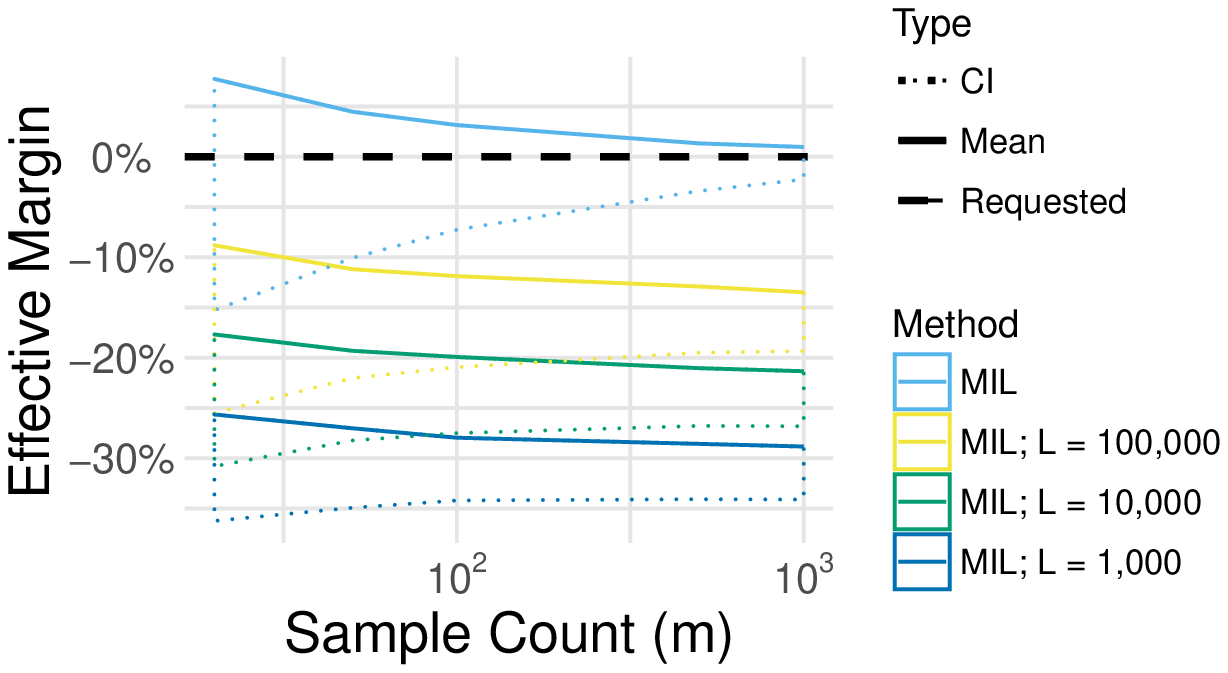}

  \caption{Comparison of approximate and exact margin in limit \TLA{} (MIL
    \TLA{}) approaches at $\cR=0.99$ (Top) and $\cR=1-10^{-7}$ (Bottom). The
    analytic approach is compared against Monte Carlo approximation using the
    delta method, varying the number $L$ of Monte Carlo samples. Note that the
    confidence bounds of the approximation converge on those of the analytic
    approach, and the approximate mean behavior is quite near the analytic
    results. Note also that in the high reliability case, the low sample count
    $L$ leads to highly under-performing designs, in terms of both mean and
    quantiles. This is due to inaccuracies in both the estimated reliability and
    margin terms.}
  \label{fig:mc_mil}
\end{figure}

\noindent Note that in order to evaluate the required moments, we formally
require the true value of $\mT$; in practice, we use a plug-in estimate. Figure
\ref{fig:mc_mil} compares the MIL \TLA{} approximation (using plug-in
estimates) against the analytic approach (using true values).

\subsection{Modeling the Margin in Probability} \label{sec:PM-margin-est}
Much like the margin in limit, we may model and approximate the margin in
probability via the delta method. The approach is nearly identical; first define
$r = R(\hat{\v\theta}) - R(\v\theta)$, and compute the partials

\begin{equation} \label{eq:MIP-gradient}
  \left.\nabla_{\hat{\v\theta}}r\right|_{\vd,\hat{\v\theta}} =
  \E_{\mX(\hat{\v\theta})}\left[\textbf{1}[g(\vd,\mX(\hat{\v\theta}))>0] %
    \frac{\nabla_{\hat{\v\theta}}\rho(\hat{\v\theta})}{\rho(\hat{\v\theta})}\right],
\end{equation}

\noindent where $\textbf{1}[\cdot]$ is the indicator function. Note that
$R(\v\theta)$ depends only indirectly upon $\hat{\v\theta}$, thus it is
eliminated in the computation of partials. The gradient above enables
first-order approximation of the moments

\begin{equation} \begin{aligned} \label{eq:MIP-moments}
    \mu_r(\vd,\hat{\v\theta}) &\approx \tilde{\mu}_r(\vd,\hat{\v\theta}) = 0, \\
    \tau_r(\vd,\hat{\v\theta})^2 &\approx \tilde{\tau}_r(\vd,\hat{\v\theta})^2 \equiv
           \left.\nabla_{\hat{\v\theta}}R\right|_{\vd,\hat{\v\theta}}^{\top}\mT_m
           \left.\nabla_{\hat{\v\theta}}R\right|_{\vd,\hat{\v\theta}},
\end{aligned} \end{equation}

\noindent which in turn enable approximation of the probability margin via

\begin{equation} \label{eq:MIP-margin}
  p \approx \Phi^{-1}(\cC)\tilde{\tau}_r(\vd,\hat{\v\theta}).
\end{equation}

\begin{figure}[!ht]
  \centering
    \includegraphics[width=0.65\textwidth]{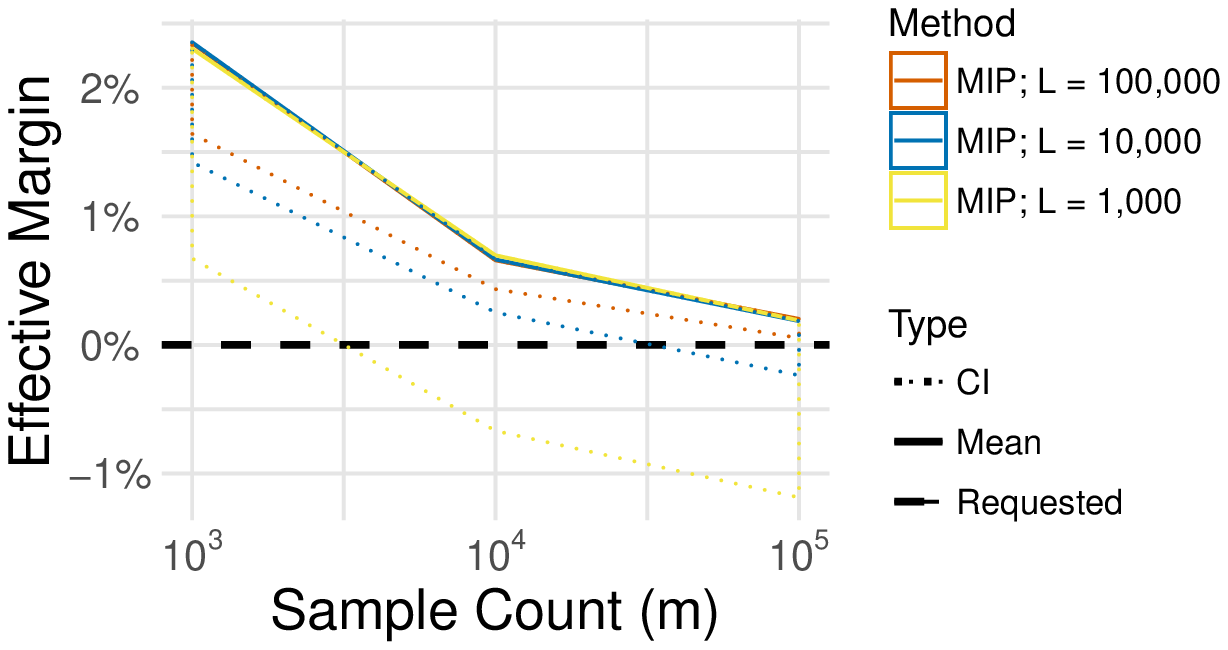} \\
    \includegraphics[width=0.65\textwidth]{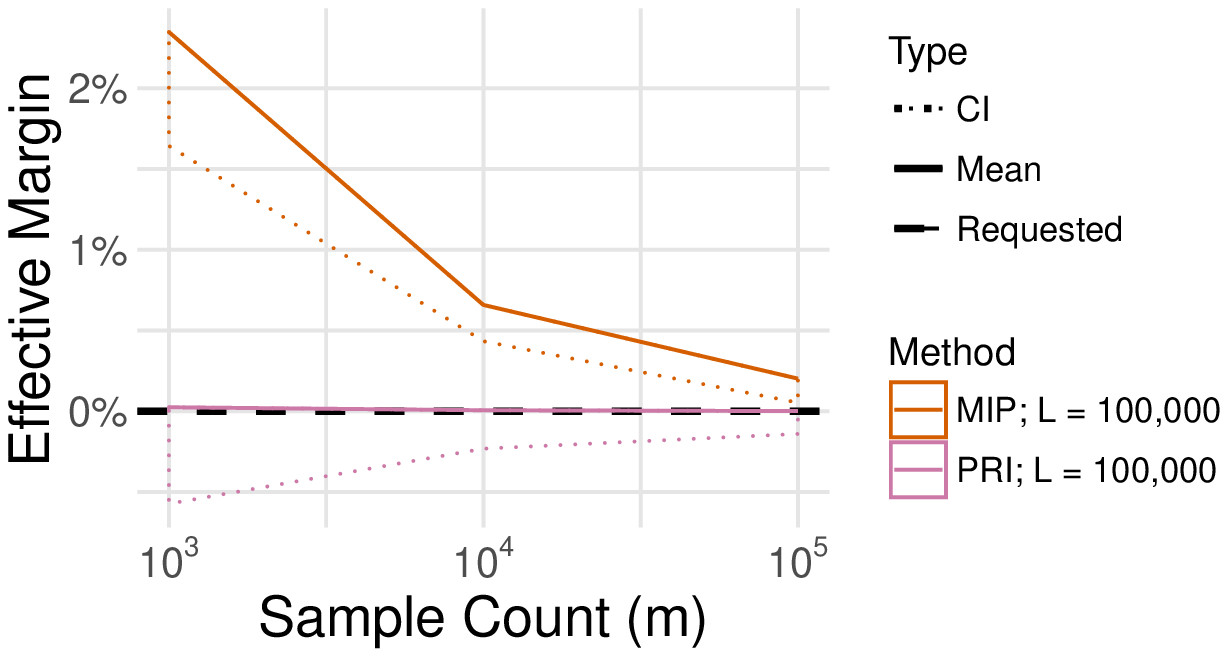}

    \caption{Effective margin for uniaxial tension at $\cR=0.90$ using
      approximate probability margin (Top), and compared against the predictive
      reliability index (PRI) approach (Bottom). Here we investigate a more lax
      reliability target, in order to illustrate an important effect: Note that
      as the sample count increases, the Monte Carlo samples ($L$) must increase
      to maintain the confidently conservative property. This implies that a
      higher-accuracy reliability calculation must be used to properly leverage
      more accurate parameter estimates. Also note that while PRI does account
      for parameter uncertainties, it is not C2, as it does not add any form of
      margin. Conspicuously, we do not present a high-reliability case
      comparison -- this is because the MIP formulation is \emph{extremely}
      expensive to run using simple Monte Carlo in the high reliability case! We
      will return to this point in Section \ref{sec:discussion}.}
  \label{fig:mc-mip}
\end{figure}

Figure \ref{fig:mc-mip} presents results for uniaxial tension using this
approximation technique within a Monte Carlo approach, compared against a
construction similar to the \textbf{predictive reliability index}
(PRI).\cite{der2008analysis} Der Kiureghian provides an approximation to the PRI
$\tilde{\beta}$ based on the delta method to the standard reliability index
$\beta(\v\theta)=\Phi^{-1}(R(\v\theta))$, given by

\begin{equation} \begin{aligned} \label{eq:PRI}
    \mu_{\beta} &\approx \Phi^{-1}(R(\hat{\v\theta})), \\
    \sigma_{\beta}^2 &= \left.\nabla_{\hat{\v\theta}}\beta\right|_{\hat{\v\theta}}^{\top}%
                \hat{\mT}\left.\nabla_{\hat{\v\theta}}\beta\right|_{\hat{\v\theta}}, \\
    \tilde{\beta} &= \frac{\mu_{\beta}}{\sqrt{1+\sigma_{\beta}^2}}.
\end{aligned} \end{equation}

\noindent Note that the PRI formally implies a Bayesian approach, while we have
so far employed frequentist constructions. Regardless, we will use the
manipulations arising from \eqref{eq:PRI} in the same fashion as the
approximations presented above, in order to provide some comparison against
existing approaches. Note that we cannot use the approximation technique of Ito
et al.,\cite{ito2018conservative} as our design problem does not take the form
of design variables perturbed by noise. One designs with the PRI via the
constraint $\tilde{\beta} \geq \Phi^{-1}(\cR)$; we present results from this
approach in Figure \ref{fig:mc-mip}. Note that \eqref{eq:PRI} effectively
inflates the variance, but provides no margin to the computed reliability index
-- Figure \ref{fig:mc-mip} demonstrates that the delta-approximated PRI behaves
much like the plug-in approach; it is not as conservative as the MIP approach.

Figure \ref{fig:mc-mip} demonstrates that careful balancing of the sample count
$m$ and number of Monte Carlo samples $n$ is necessary to approach the C2
property promised by the analytic MIP approach. We perform a scalar analysis
(Appendix \ref{sec:balancing-error}) to study this phenomenon, and find that the
Monte Carlo estimated variance $\hat{\tau}^2$ has variance in excess of
$\tilde{\tau}^2$ approximated by

\begin{equation} \label{eq:var-balance}
  \V[\hat{\tau}^2] \approx \V[\tilde{\tau}^2]\left(1 + k\frac{m}{n}\right),
\end{equation}

\noindent where $m$ is the sample count, $n$ is the number of Monte Carlo
samples, and $k\in\R{}_{>0}$ is an unknown constant. \Cref{eq:var-balance}
illustrates that the estimated margin $\hat{p_{\cC}} = \Phi^{-1}(\cC)\hat{\tau}$
has dispersion in excess of that considered in the delta method. An increase in
$m$ must be met with a comparable increase in $n$, in order to combat this
deleterious effect.

\subsection{Integration and Implementation} \label{sec:integration}
Before moving on to our final example, we first discuss the practical
integration of \TLA{} into a reliability-based design optimization (RBDO)
procedure. In order to fully realize the efficiency promised by the
approximation techniques above, particular integration choices must be made when
implementing the design and analysis loops.

First, since \TLA{} may (in general) depend on the design variables $\vd$, it
must be estimated alongside the system reliability. In both the margin in limit
and margin in probability approaches we provide $\hat{\v\theta},\hat{\mT}$,
select $\cR,\cC$, and enforce a modified constraint. In the margin in limit
approach, we enforce

\begin{equation} \begin{aligned} \label{eq:MIL-constraint}
  \P_{\mX(\hat{\v\theta})}[g(\vd,\mX(\hat{\v\theta})) > g_{MIL,\cC}] &\geq \cR, \\
  \P_{\mX(\hat{\v\theta})}\left[g_{MIL,\cC} > \E_{\mX(\hat{\v\theta})}[g(\vd,\mX(\hat{\v\theta}))]%
         - \E_{\mX}[g(\vd,\mX)]\right] &= \cC,
\end{aligned} \end{equation}

\noindent while in the margin in probability approach, we enforce

\begin{equation} \begin{aligned} \label{eq:MIP-constraint}
  \P_{\mX(\hat{\v\theta})}[g(\vd,\mX(\hat{\v\theta})) > 0] &\geq \cR + p, \\
  \P_{\hat{\v\theta}}\left[R(\hat{\v\theta}) > \cR + p\right] &= \cC.
\end{aligned} \end{equation}

\noindent In the case where the reliability analysis is nested within an
optimization loop, the approach is called bi-level;\cite{eldred2006second} --
confusingly, some authors refer to this nesting as a `double loop'. For clarity,
we note that in this work we seek to address the statistical double loop; other
authors have addressed the bi-level issue.\cite{agarwal2004unilevel} Within a
particular reliability analysis at value $\vd$, we first obtain realizations of
the limit state $g(\vd,\mX_i(\hat{\v\theta}))$, either directly
(non-intrusively) or by sampling a constructed surrogate (e.g. via an intrusive
procedure). We then use these realizations to compute the margin of choice,
which we then apply to the reliability problem. Algorithm \ref{alg:margin}
illustrates both the margin in limit and margin in probability procedures in
pseudocode, using simple Monte Carlo.

\begin{algorithm}[!ht]
\begin{minipage}{0.45\textwidth}
  \KwData{$\hat{\v\theta},\hat{\mT}_m$;$\cR,\cC,\epsilon,n$}
  \KwResult{$\vd^*$}
  Select $\vd_0$ \\
  \While{$C(\vd_j)$ not converged within $\epsilon$}{
    Reliability Analysis \\
    \For{$i = 1:n$}{
      $\mX_i \sim \rho(\hat{\v\theta})$ \\
      $g_i \gets g(\vd_j,\mX_i)$ \\
      $\nabla_{\hat{\v\theta}} D_i \gets \frac{g_i-\overline{g}}{\rho(\mX_i;\hat{\v\theta})}
        \nabla_{\hat{\v\theta}} \rho(\mX_i;\hat{\v\theta})$ \\
    }

    \bigskip
    Margin Computation \\
    $\nabla_{\hat{\v\theta}} D \gets \frac{1}{n}\sum_{i=1}^n \nabla_{\hat{\v\theta}} D_i$ \\
    $\tau_D^2 \gets \nabla_{\hat{\v\theta}} D^{\top} \hat{\mT}_m \nabla_{\hat{\v\theta}} D$ \\
    $g_{MIL,\cC} = \Phi^{-1}(\cC)\sqrt{\tau_D^2}$ \\
    $R(\hat{\v\theta}) \gets \frac{1}{n}\sum_{i=1}^n \textbf{1}[g_i-g_{MIL,\cC}>0]$ \\

    \bigskip
    Design Optimization \\
    \textbf{Select} $\vd_{j+1}$ \textbf{such that} \\
    \textbf{Minimize} $C(\vd_{j+1})$ \\
    \textbf{Subject to} $R(\hat{\v\theta})\geq \cR$ \\

    \bigskip
    Iterate \\
    $j\gets j+1$
  }
  \Return $\vd^* \gets \vd_j$
\end{minipage} %
\begin{minipage}{0.45\textwidth}
  \KwData{$\hat{\v\theta},\hat{\mT}_m$;$\cR,\cC,\epsilon,n$}
  \KwResult{$\vd^*$}
  Select $\vd_0$ \\
  \While{$C(\vd_j)$ not converged within $\epsilon$}{
    Reliability Analysis \\
    \For{$i = 1:n$}{
      $\mX_i \sim \rho(\hat{\v\theta})$ \\
      $g_i \gets g(\vd_j,\mX_i)$ \\
      $\nabla_{\hat{\v\theta}} R_i \gets \frac{\textbf{1}[g_i>0]}%
                                {\rho(\mX_i;\hat{\v\theta})}
                                \nabla_{\hat{\v\theta}} \rho(\mX_i;\hat{\v\theta})$ \\
    }

    \bigskip
    Margin Computation \\
    $\nabla_{\hat{\v\theta}} R \gets \frac{1}{n}\sum_{i=1}^n \nabla_{\hat{\v\theta}} R_i$ \\
    $\tau_R^2 \gets \nabla_{\hat{\v\theta}} R^{\top} \hat{\mT}_m \nabla_{\hat{\v\theta}} R$ \\
    $p_{\cC} = \Phi^{-1}(\cC)\sqrt{\tau_R^2}$ \\
    $R(\hat{\v\theta}) \gets \frac{1}{n}\sum_{i=1}^n \textbf{1}[g_i>0]$ \\

    \bigskip
    Design Optimization \\
    \textbf{Select} $\vd_{j+1}$ \textbf{such that} \\
    \textbf{Minimize} $C(\vd_{j+1})$ \\
    \textbf{Subject to} $R(\hat{\v\theta})\geq \cR + p_{\cC}$ \\

    \bigskip
    Iterate \\
    $j\gets j+1$
  }
  \Return $\vd^* \gets \vd_j$
\end{minipage}

  \bigskip
  \caption{Performing reliability-based design optimization with margin in limit
    (Left) and probability (Right), using simple Monte Carlo. Here
    $\textbf{1}[\cdot]$ denotes the indicator function. For brevity, the sample
    mean $\overline{g}$ is used before it is formally available. The
    optimization algorithm employed is purposefully not specified to emphasize
    the generality of the margin algorithms. Note that in both implementations,
    the margin computation uses information already available from the
    reliability analysis. Since the evaluation of the limit state $g_i$ is
    commonly the most expensive portion of the analysis, the computation of
    margin in these approaches adds negligible computational expense.}
  \label{alg:margin}
\end{algorithm}

\FloatBarrier
\section{Demonstration: Cantilevered Beam} \label{sec:demonstration}

\begin{figure}[!ht]
  \centering
  \includegraphics[]{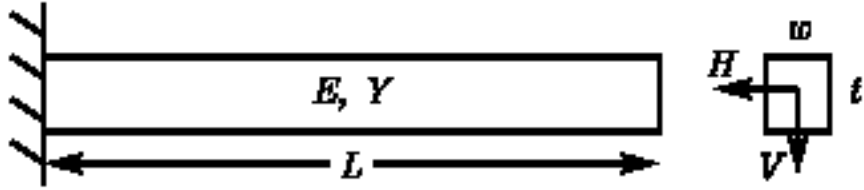}
  \caption{Schematic for the proposed cantilever beam problem subject to a
    lateral and vertical load \cite{eldred2007investigation}. Material
    properties ($E,Y$) and loading ($H,V$) are uncertain. $L=100$ inches and $w$
    and $t$ are deterministic design variables. }
  \label{fig:cantilever_beam}
\end{figure}

As a demonstration of the application of \margin{} in a reliability-based design
optimization, we consider the design of a cantilevered beam
\cite{eldred2007investigation}. Figure \ref{fig:cantilever_beam} illustrates the
problem of a rectangular constant cross-section cantilevered beam subject to a
lateral load $H$ and vertical load $V$ at its end. Both loads and the beam's
elastic modulus $E$ and yield strength $Y$ are assumed to be normally
distributed as shown in table \ref{tab:beam_random_variables}; thus the problem
has 4 random variables $\mathbf{X} = [H, V, E, Y]^{\top}$. For this problem, we
consider exact knowledge of the load distributions, but estimate distribution
parameters for material properties $E$ and $Y$ via sampling. The designer has
control over two deterministic variables $\vd = [w, t]^{\top}$, the width $w$
and the thickness $t$ of the beam. The quantities of interest for this problem
include the cross-sectional area of the beam $wt$, as well as the stress and
displacement of the beam, which are desired to not exceed the yield strength $Y$
and maximum allowable displacement $D_0 = 2.2535$ inches of the beam.

\begin{table}[!ht]
\caption{Truth distributions for the random variables in the cantilevered beam
  problem.}
\label{tab:beam_random_variables}
\begin{center}
\begin{tabular}[]{ c | c | c }
Name & Variable & Distribution \\ \hline
Lateral load & $H$ & $\mathcal{N}(500, 100^2)$ \\ \hline
Vertical load & $V$ & $\mathcal{N}(1000, 100^2)$ \\ \hline
Elastic modulus & $E$ & $\mathcal{N}(2.9\e{7}, (1.45\e{6})^2)$ \\ \hline
Yield strength & $Y$ & $\mathcal{N}(40000, 2000^2)$ \\ \hline
\end{tabular}
\end{center}
\end{table}

The stress $S$, and the displacement $D$ in the beam are given by:

\begin{equation}
  S(\vd, \mX) = \frac{600V}{wt^2} + \frac{600H}{w^2t},
\end{equation}

\begin{equation}
  D(\vd, \mX) = \frac{4L^3}{Ewt} \sqrt{ \left( \frac{V}{t^2} \right)^2 + %
              \left( \frac{H}{w^2} \right)^2},
\end{equation}

\noindent whereupon the normalized limit state functions $g_S$ and $g_D$ are
written as:

\begin{equation}
  g_S(\vd, \mX) = 1 - \frac{S(\vd,\mX)}{Y},
\end{equation}

\begin{equation}
  g_D(\vd, \mX) = 1 - \frac{D(\vd,\mX)}{D_0},
\end{equation}

\noindent In the MIL implementation, we then formulate and solve the following
minimum cross-sectional area (\textit{i.e.} minimum mass) design problem with
chance constraints for the probability of failure to not exceed 0.135\%:

\begin{equation} \begin{aligned} \label{eq:beam-MIL-optimization}
    \text{min }  & C(\vd) = wt, \\
    \text{s.t. } & R_S(\hat{\v\theta})\equiv\P_{\mX(\hat{\v\theta})}%
         [g_S(\vd,\mX(\hat{\v\theta})) > g_{S,MIL,\cC}] \geq \cR = 0.99865, \\
                 & R_D(\hat{\v\theta})\equiv\P_{\mX(\hat{\v\theta})}%
         [g_D(\vd,\mX(\hat{\v\theta})) > g_{D,MIL,\cC}] \geq \cR = 0.99865, \\
                 & 1 \leq w,t \leq 4 \\
\end{aligned} \end{equation}

\noindent where the limit state margins $g_{S,MIL,\cC}$ and $g_{D,MIL,\cC}$
defined by the equality constraints in equation \ref{eq:MIL-constraint}
calculated for confidence interval $\cC = 0.95$ are implicit. Similar
manipulations yield the MIP approach. In practice, we reformulate the
constraints using the performance measure approach by rewriting them using the
inverse CDF of the limit state functions \cite{tu1999pma}. Such a formulation
avoids issues during gradient-based optimization when the calculated reliability
is 100\%.

\begin{figure}[!ht]
  \centering
  \includegraphics[width=0.65\textwidth]{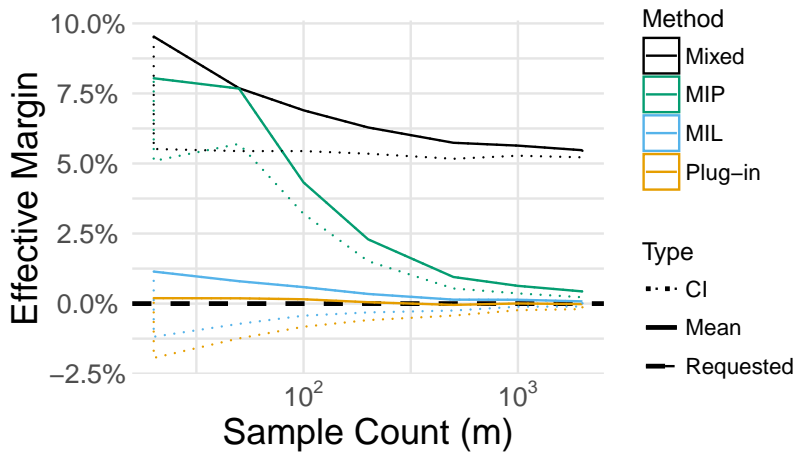}

  \caption{Effective margin for optimal cantilever beam designs using the
    plug-in approach (PI), basis value approach, and proposed margin in limit
    (MIL) and margin in probability (MIP) \margin{} approaches. Confidence
    intervals are constructed via a normal approximation from 40 independent
    optimization results, and chance constraints are estimated via Monte Carlo
    sampling with $N=1\e{5}$ samples. Note the excessive margin when an A-basis
    value is used, and improved margin over the plug-in approach when the \TLA{}
    approaches are employed, especially for the margin in probability approach.
    Note also that the approximate MIP approach is \emph{not} C2 in this case,
    as it is somewhat over-conservative, but does approach the theoretical
    ideal.}
  \label{fig:beam_results}
\end{figure}

\begin{figure}[!ht]
  \centering
    \includegraphics[width=0.65\textwidth]{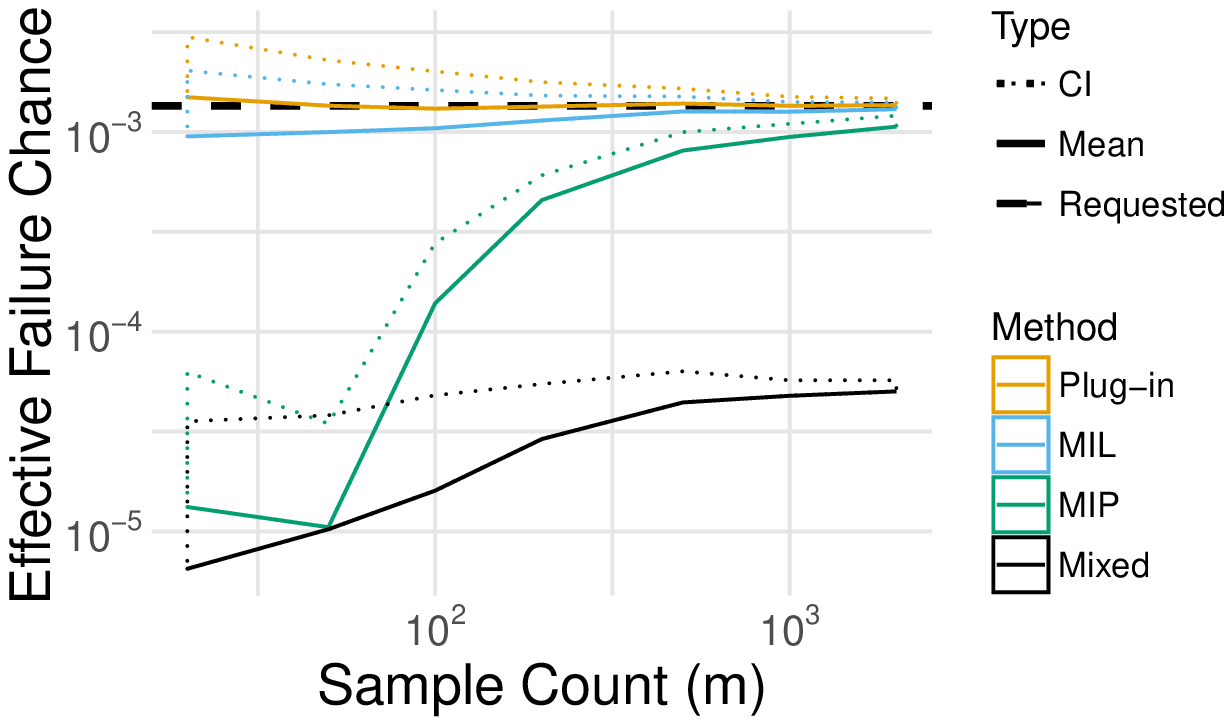} \\
    \includegraphics[width=0.65\textwidth]{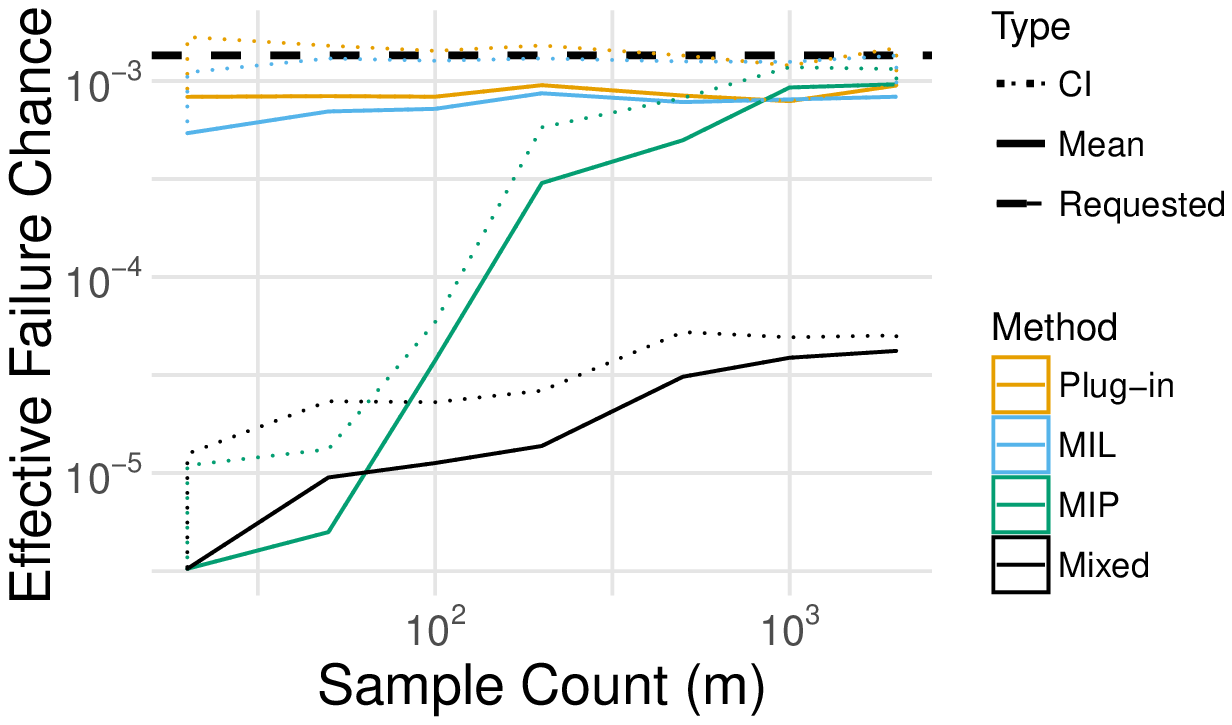}

    \caption{Effective reliabilities with respect to stress (Top) and
      displacement (Bottom) constraints corresponding to the optimization
      results shown in figure \ref{fig:beam_results}. Optimization with basis
      values leads to an overly conservative design, whereas design with the
      \margin{} leads to designs with performance closer to the requested
      performance. In particular, the margin in probability (MIP) \TLA{}
      approach leads to satisfactorily conservative reliabilities which approach
      the design goal as information about material properties increases.}
  \label{fig:beam_results_reliabilities}
\end{figure}

We compare the results of the reliability-based optimization problem in Figures
\ref{fig:beam_results} and \ref{fig:beam_results_reliabilities}, and Tables
\ref{tab:beam_results_m100} and \ref{tab:beam_results_m1000} for several
methods: the plug-in approach, mixed approach, and proposed \margin{}
implementations. We first observe that optimization using the plugin approach
leads to designs with an unbiased effective margin which, on average, satisfies
the reliability constraints. On the other hand, optimization with basis values
leads to excessively conservative designs for this problem, with a large
effective margin and extremely high reliability far from the desired value.

In contrast, optimization with the proposed \margin{} approaches leads to
conservative designs which have positive effective margin and trend towards the
desired design reliability with increasing sample count. In particular the
delta-approximated margin in probability (MIP) implementation of the \margin{}
is desirable from an engineering perspective: Although it is not C2 (it is
over-conservative), it leads to conservative designs when little information is
available about material properties. The approximate MIP approach is able to
capitalize on improved information (greater $m$), and approaches true C2
behavior. A comparison of the results in tables \ref{tab:beam_results_m100} and
\ref{tab:beam_results_m1000} also illustrates the increased effectiveness of the
MIP implementation at higher sample counts leading to less conservative but
reliable designs.

\begin{table}[!ht]
\caption{Comparison of cantilever beam optimization results for a sample count
  of 100 (\textit{i.e.} using information from 100 material property
  characterization tests). The average objective (cross-sectional area) value
  and constraint reliabilities are shown with coefficients of variation in
  parentheses. For a constant density beam, average weight savings of 2.4\% are
  realized for design using the margin in probability (MIP) \TLA{} approach
  compared to design using basis values (BV). Design using the margin in limit
  (MIL) \TLA{} approach yields average weight savings of 5.9\%.}
\label{tab:beam_results_m100}
\begin{center}
\begin{tabular}[]{ c | c | c | c | c | c }
  Method & Objective & $\E[t]$ & $\E[w]$ & $R_S$ & $R_D$ \\ \hline
  MC+PI      & 9.53 $(6.02\e{-3})$   & 3.82 & 2.49 & 0.99869 $(4.34\e{-4})$ & 0.99913 $(3.47\e{-4})$ \\ \hline
  MC+BV      & 10.17 $(8.38\e{-3})$  & 3.90 & 2.61 & 0.99998 $(1.98\e{-5})$ & 0.99999 $(7.23\e{-6})$ \\ \hline
  MC+MIL PM & 9.57 $(6.20\e{-3})$   & 3.84 & 2.49 & 0.99896 $(3.40\e{-4})$ & 0.99924 $(3.28\e{-4})$ \\ \hline
  MC+MIP PM & 9.93 $(6.63\e{-3})$   & 3.88 & 2.56 & 0.99986 $(8.54\e{-5})$ & 0.99996 $(1.31\e{-5})$ \\ \hline
\end{tabular}
\end{center}
\end{table}

\begin{table}[!ht]
\caption{Comparison of cantilever beam optimization results for a sample count
  of 1000 (\textit{i.e.} using information from 1000 material property
  characterization tests). The average objective (cross-sectional area) value
  and constraint reliabilities are shown with coefficients of variation in
  parentheses. For a constant density beam, average weight savings of 4.7\% are
  realized for design using the margin in probability (MIP) \TLA{} approach
  compared to design using basis values (BV). Note the increased savings compared to when a sample count of 100 is used. Design using the margin in limit
  (MIL) \TLA{} approach yields average weight savings of 5.2\%.}
\label{tab:beam_results_m1000}
\begin{center}
\begin{tabular}[]{ c | c | c | c | c | c }
  Method & Objective & $\E[t]$ & $\E[w]$ & $R_S$ & $R_D$ \\ \hline
  MC+PI      & 9.51 $(1.47\e{-3})$   & 3.79 & 2.51 & 0.99865 $(9.11\e{-5})$ & 0.99926 $(1.85\e{-4})$ \\ \hline
  MC+BV      & 10.05 $(2.09\e{-3})$  & 4.00 & 2.51 & 0.99995 $(5.77\e{-6})$ & 0.99996 $(6.75\e{-6})$ \\ \hline
  MC+MIL PM & 9.53 $(1.51\e{-3})$   & 3.81 & 2.50 & 0.99874 $(9.11\e{-5})$ & 0.99920 $(2.75\e{-4})$ \\ \hline
  MC+MIP PM & 9.58 $(1.69\e{-3})$   & 3.87 & 2.48 & 0.99906 $(9.56\e{-5})$ & 0.99913 $(2.63\e{-4})$ \\ \hline
\end{tabular}
\end{center}
\end{table}

\section{Discussion} \label{sec:discussion}
In this work, we introduced the concept of \textbf{\margin{}} to aid in
addressing statistical uncertainties in RBDO. \TLA{} is, by construction,
capable of controlling a system limit state and avoiding excessive design
conservatism. To show the flexibility of this concept, we introduced two
operationalizations of the \TLA{} concept, introducing margin in limit, and
margin in probability. The latter provided an additional benefit: the ability to
guarantee a desired reliability at a designed confidence level, what we called
confidently conservative (C2). We derived an approximation for and demonstrated
the efficacy of both approaches on a classic reliability test case -- the
cantilever beam problem -- which reduced excess weight by $2-5\%$ when compared
with design using an A-basis value, while maintaining the desired reliability at
(or above) the desired confidence level. This demonstrates the potential of MIP
and other \TLA{} strategies to produce tangible gains in engineering design for
reliability.

Practically, what must be done to perform design for reliability using \TLA{}
instead of basis values? For the MIP approach suggested above, one must first
model the material properties with random variables, in line with existing
military design guidelines.\cite{mil-hdbk-17-3e1997} One must then estimate the
parameters $\hat{\v\theta}$ for these random variables -- this is already done
in some approaches to computing basis values.\cite{barbero2012} In addition, one
must estimate a covariance matrix $\hat{T}_m$ for the estimated parameters, for
use in the delta method. Finally, one must perform RBDO with MIP as illustrated
above; the computational expense of this effort will scale with the desired
reliability tolerance, and with the cost involved with system simulation.

Of course, further efforts are necessary to develop and deploy the \TLA{}
concept. Numerous algorithms and software packages for design for reliability
exist, which could benefit from integration with a \TLA{} implementation.
Acceleration is also key; in this work we considered simple Monte Carlo, which
is known to be slow to converge -- concretely, this stymied our efforts to apply
MIP in the high-reliability case. Integrating \TLA{} with fast integrators and
quadrature rules is a clear next step. The current implementations of \TLA{}
suggested here lean heavily on an assumed distribution; this weakness could be
lessened by using a more general random variable model, such as the Johnson
distribution.\cite{mcdonald2013probabilistic} Furthermore, it would be desirable
to have a non-parametric (empirical) way to implement \TLA{} -- an approach
which would ideally be robust to departures from modeled randomness. An
application of the \textbf{ambiguity set} may aid in non-parameteric
efforts.\cite{kapteyn2018ambiguity-set} Operationally, it may be beneficial to
formulate both the design and sampling plan within the same optimization, using
margin as a link -- recent developments in multi-objective optimization
leveraging stochastic dominance appear to be an attractive path
forward.\cite{cook2018stochastic-dominance} Finally, we reiterate that \TLA{} is
intended to cover statistical uncertainties only -- uncertainties addressed by
Factors of Safety include unknown unknowns, so a \TLA{} could never replace a
FOS. However, we believe that \margin{} is an early but key component of
quantifying, propagating, and above all \emph{managing} uncertainty in
engineering design.

\section*{Acknowledgments}
This work sprung from numerous conversations the first author had while working
as an intern at the Northop-Grumman Corporation; thus thanks are owed to many
NGC engineers. In particular, he would like to thank John Madsen, who served as
an incredible mentor at NGC on professional, technical, and personal
development.

The first author was supported in part by the NSF GRFP under Grant No.
DGE-114747. The first two authors would like to acknowledge the support of the
DARPA Enabling Quantification of Uncertainty in Physical Systems (EQUiPS)
program.

\bibliographystyle{ama}
\bibliography{main}

\section{Appendix}

\subsection{Margin in Limit \TLA{}} \label{sec:MIL-pf}
In this entry, we prove that the margin in limit (MIL) approach satisfies
convergence property \ref{def:sm2}. This Appendix entry adopts the more standard
statistics notation of Van der Vaart\cite{van1998asymptotic}, in contrast with
the bulk of the manuscript.

\noindent\underline{Claim:} Suppose
$\sqrt{m}(\hat{\v\theta}-\v\theta)\stackrel{d}{\to} \dN(0,\mT)$. Then the margin
$g_{MIL,\cC}$ defined in \eqref{eq:MIL-def} is asymptotically zero, satisfying
Point \ref{def:sm2}.

\bigskip
\noindent\underline{Pf:} Define

\begin{equation}
  D(\vd,\hat{\v\theta}) = E_{\mX(\hat{\v\theta})}[g(\vd,\mX(\hat{\v\theta}))] -
  \E_{\mX}[g(\vd,\mX(\v\theta)),
\end{equation}

\noindent an application of the delta method \cite{van1998asymptotic} yields

\begin{equation}
  \sqrt{m}(D(\vd,\hat{\v\theta})-D(\vd,\v\theta))\stackrel{d}{\to}
  \dN(0,\tau'^2),
\end{equation}

\noindent where
$\tau'^2=\left.\nabla_{\v\theta}D\right|_{\vd,\v\theta}^{\top}\mT\left.\nabla_{\v\theta}D\right|_{\vd,\v\theta}$.
Note that $D(\vd,\v\theta)=0$. By the definition of $g_{MIL,\cC}$, we have the limit

\begin{equation}
  g_{MIL,\cC} \to \frac{\tau'}{\sqrt{m}}\Phi^{-1}(C),
\end{equation}

\noindent which completes the proof.$\square$

\subsection{Margin in Probability \TLA{}} \label{sec:MIP-pf}
In this entry, we prove that the margin in probability (MIP) approach satisfies
convergence property \ref{def:sm2}.

\noindent\underline{Claim:} Suppose
$\sqrt{m}(\hat{\v\theta}-\v\theta)\stackrel{d}{\to} \dN(0,\mT)$. Then the margin
$p_{\cC}$ defined in \eqref{eq:MIL-def} is asymptotically zero, satisfying
Point \ref{def:sm2}.

\bigskip
\noindent\underline{Pf:} Define

\begin{equation}
  r(\vd,\hat{\v\theta}) = R(\vd,\hat{\v\theta}) - R(\vd,\hat{\v\theta}),
\end{equation}

\noindent an application of the delta method \cite{van1998asymptotic} yields

\begin{equation}
  \sqrt{m}(r(\vd,\hat{\v\theta}) - r(\vd,\v\theta))\stackrel{d}{\to}
  \dN(0,\tau'^2),
\end{equation}

\noindent where
$\tau'^2=\left.\nabla_{\v\theta}r\right|_{\vd,\v\theta}^{\top}\mT\left.\nabla_{\v\theta}r\right|_{\vd,\v\theta}$.
Note that $r(\vd,\v\theta)=0$. By the definition of $p_{\cC}$, we have the limit

\begin{equation}
  p_{\cC} \to \frac{\tau'}{\sqrt{m}}\Phi^{-1}(C),
\end{equation}

\noindent which completes the proof.$\square$

\subsection{Conservative Reliability Index \TLA{}} \label{sec:CRI-pf}
In this entry, we prove that the conservative reliability index (CRI) satisfies
convergence property \ref{def:sm2}.

\noindent\underline{Claim:} Suppose
$\sqrt{m}(\hat{\v\theta}-\v\theta)\stackrel{d}{\to} \dN(0,\mT)$, and that
$R(\hat{\v\theta})$ is differentiable with respect to $\v\theta$ at the true
parameter value. Then the quantile $R^{\alpha}\to R(\v\theta)$, satisfying Point
\ref{def:sm2}.

\bigskip
\noindent\underline{Pf:} Recall

\begin{equation} \label{eq:cri-aux}
  \P_{\hat{\theta}}[R(\hat{\v\theta}) - R^{\alpha} > 0] = \alpha,
\end{equation}

\noindent an application of the delta method \cite{van1998asymptotic} to
the approximate reliability yields

\begin{equation}
  \sqrt{m}(R(\hat{\v\theta}) - R(\v\theta)) \stackrel{d}{\to} \dN(0, \tau'^2),
\end{equation}

\noindent where
$\tau'^2=\left.\nabla_{\v\theta}R\right|_{\vd,\v\theta}^{\top}\mT\left.\nabla_{\v\theta}R\right|_{\vd,\v\theta}$.
The asymptotic solution to auxillary equation \eqref{eq:cri-aux} is then given
by

\begin{equation}
  R^{\alpha} \to R(\v\theta) - \frac{\tau'}{\sqrt{m}}\Phi^{-1}(\alpha),
\end{equation}

\noindent thus the constraint $R^{\alpha}\geq\cR$ recovers the true reliability
constraint $R(\v\theta)\geq\cR$ in the asymptotic limit. $\square$

\subsection{Balancing Error} \label{sec:balancing-error}
In this entry, we study the error contributions to margin in probability using a
delta approximation. Suppose we have a single parameter $\theta$, estimated by
$t_m$, and wish to apply margin $p_{\cC}$ of the form

\begin{equation} \begin{aligned}
    \hat{p}_{\cC}       &= \Phi^{-1}(\cC)\hat{\tau}, \\
    \hat{\tau}^2 &= \hat{r'}t_m\hat{r'},
\end{aligned} \end{equation}

\noindent where $r' = \left.\frac{dr}{d\theta}\right|_{t_m}$, and $\hat{r'}$ is
a Monte Carlo approximation of the type described in Section
\ref{sec:gradient-trick}. Note that $\hat{\tau}^2$ is an estimate; thus it is
itself randomly distributed. While the delta method guarantees convergence for
the estimate $\tilde{\tau}$, it does not account for additional variability
arising from the Monte Carlo estimate of the derivative. The following
investigation considers the effects of Monte Carlo on $\hat{\tau}^2$, compared
with $\tilde{\tau}^2 = t_mr'^2$.

The central limit theorem endorses the following random variable models

\begin{equation} \begin{aligned}
    t_m      &\sim \dN(\theta, \gamma^2/m), \\
    \hat{r'} &\sim \dN(r', \sigma^2/n).
\end{aligned} \end{equation}

\noindent One may show that $\hat{\tau}^2$ has moments given by

\begin{equation} \begin{aligned}
    \E[\hat{\tau}^2] &= \theta(r'^2 + \sigma^2/n), \\
    \V[\hat{\tau}^2] &= \frac{\gamma^2}{m}r'^4\left[
      1 + \left(\frac{3}{r'^4} + 2\frac{\theta^2m}{r'^4\gamma^2}\right)
      \left(2\frac{\sigma^2}{n}r'^2 + \frac{\sigma^4}{n^2}\right)\right].
\end{aligned} \end{equation}

\noindent The expectation $\E[\hat{\tau}^2]$ illustrates bias in our estimate,
but it is always positive, and thus only increases conservatism. In the limit
$n,m>>1$, we can approximate

\begin{equation} \begin{aligned} \label{eq:excess-variance}
  \V[\hat{\tau}^2] &\approx \frac{\gamma^2}{m}r'^4
    \left[1 + 4\frac{\theta^2\sigma^2}{\gamma^2r'^4}\frac{m}{n}\right], \\
                   &= \V[\tilde{\tau}^2]\left(1 + k\frac{m}{n}\right).
\end{aligned} \end{equation}

\noindent \Cref{eq:excess-variance} enables us to understand the error
properties demonstrated in Figure \ref{fig:mc-mip}: Compared with the quantity
$\tilde{\tau}^2$ considered in the delta method, the estimate $\hat{\tau}^2$ has
excess variance, scaled by the factor $k\frac{m}{n}$. If the sample count $m$ is
increased without a comparable increase in Monte Carlo samples $n$, then the
excess variance can result in both over- and under-estimated margin terms. It is
these under-estimated cases that foil the C2 property of the estimated MIP
procedure.

While \eqref{eq:excess-variance} suggests that `balancing' the sample count $m$
and Monte Carlo samples $n$ is desirable, one cannot make a more precise
statement without knowing the value of the constant $k$, which in general will
be unknown. A practical heuristic is to seek $m << n$; fortunately, physical
samples $m$ will often be considerably more expensive to gather than
computational samples $n$.

\subsection{Example: Bias in Reliability Calculation} \label{sec:ex-bias}
As mentioned above, the CRI framework of Ito et al.\cite{ito2018conservative} is
attractive, but susceptible to bias. Here we provide an example RBDO problem
which illustrates this issue. We consider the following problem

\begin{equation} \begin{aligned} \label{eq:ex-bias}
    \text{min. } &d, \\
    \text{s.t. } &\P_{X(\lambda)}[d - X \geq 0] \geq \cR, \\
                 &d \geq 0,
\end{aligned} \end{equation}

\noindent where $\lambda$ parameterizes an exponential random variable $X \sim
\text{exp}(1)/\lambda$. In this case $R(\lambda) = 1 - exp(-\lambda d)$, so $d^*
= -\log(1 - \cR)/\lambda$. Our objective in this example is to approximate a
solution to RBDO problem \eqref{eq:ex-bias}, in the absence of the true value of
$\lambda$, but given samples from the true distribution $X_i \sim
\text{exp}(1)/\lambda$. The maximum likelihood estimator for $\lambda$ is given
by

\begin{equation}
  \hat{\lambda} = 1 / \overline{X},
\end{equation}

\noindent which is known to be biased estimator. While we can easily
re-parameterize the exponential distribution to avoid this issue, more generally
one may want to work with biased estimators, for instance to take advantage of
Stein's phenomenon.\cite{efron1977stein}

The original Ito et al. work is framed in terms of failure probabilities, so we
carry out the trivial transform to this form, seeking a desired failure
probability $\cF = 1 - \cR$, and consider the CRI approach

\begin{equation} \begin{aligned} \label{eq:ex-bias-cri}
    \text{min. } &d, \\
    \text{s.t. } &F^{\alpha} \leq \cF, \\
                 &d \geq 0, \\
                 &\P_{\hat{\lambda}}[F(\hat{\lambda}) < F^{\alpha}] = \alpha,
\end{aligned} \end{equation}

\noindent in comparison with the MIP approach

\begin{equation} \begin{aligned} \label{eq:ex-bias-mip}
    \text{min. } &d, \\
    \text{s.t. } &F(\hat{\lambda}) + p \leq \cF, \\
                 &d \geq 0, \\
                 &\P_{\hat{\lambda}}[F(\lambda) - F(\hat{\lambda}) < p] = \alpha.
\end{aligned} \end{equation}

\noindent Practically, we cannot use the approximation technique suggested in
Ito et al.\cite{ito2018conservative}, as the shape of the sampling distribution
for $\hat{\lambda}$ is dependent on the design variables. We solve both
optimization problems semi-analytically, using a monte-carlo approximation for
the sampling distribution of $\hat{\lambda}$. We report the single result
arising from \eqref{eq:ex-bias-cri}, along with the $\alpha$ -percentile case
from \eqref{eq:ex-bias-mip} in Table \ref{tab:ex-bias-res}.

\begin{table}[!ht]
  \caption{Effective margin results from example RBDO problems posed in
    \eqref{eq:ex-bias-cri} and \eqref{eq:ex-bias-mip}, with $\cF = 0.01, \alpha
    = 0.9$, reported against sample count $m$. Note that the CRI approach has an
    effective margin of nearly $-50\%$ at low sample counts; this is due to the
    unhandled bias entering through the exponential parameter estimator. The MIP
    approach achieves an effective margin near machine precision $\epsilon$.}
  \label{tab:ex-bias-res}
  \begin{tabular}{@{}l|lllllll@{}}
    $m$ & 5 & 10 & 25 & 50 & 100 & 500 & 1000 \\
    $M_{eff,CRI}$ & $-0.496$ & $-0.376$ & $-0.239$ & $-0.176$ & $-0.129$ & $-0.058$ & $-0.043$ \\
    $M_{eff,MIP}$ & $O(\epsilon)$ & $O(\epsilon)$ & $O(\epsilon)$ & $O(\epsilon)$ & $O(\epsilon)$ & $O(\epsilon)$ & $O(\epsilon)$
  \end{tabular}
\end{table}

\end{document}